\documentclass{amsart}
\usepackage{amssymb}

\usepackage{amscd}
\usepackage{amsmath}
\usepackage{graphicx}

\newtheorem{theorem}{Theorem}
\theoremstyle{plain}

\newtheorem{definition}{Definition}

\newtheorem{proposition}{Proposition}

\numberwithin{equation}{section}

\input{tcilatex}

\begin{document}
\title{ENTANGLEMENT, QUANTUM\ ENTROPY AND MUTUAL\ INFORMATION}
\author{Viacheslav P Belavkin}
\address{Department of Mathematics, University of Nottingham, NG7 2RD Nottingham, UK}
\author{Masanori Ohya}
\address{Department of Information Sciences, Science University of Tokyo, 278 Noda
City, Chiba, Japan}
\thanks{The authors acknowledge the support under the JSPS Senior Fellowship Program
and the Royal Society scheme for UK-Japan research collaboration.}
\date{July 20, 1998}
\subjclass{Quantum Probability and Information}
\maketitle

\begin{abstract}
The operational structure of quantum couplings and entanglements is studied
and classified for semifinite von Neumann algebras. We show that the
classical-quantum correspondences such as quantum encodings can be treated
as diagonal semi-classical (d-) couplings, and the entanglements
characterized by truly quantum (q-) couplings, can be regarded as truly
quantum encodings. The relative entropy of the d-compound and entangled
states leads to two different types of entropy for a given quantum state:
the von Neumann entropy, which is achieved as the maximum of mutual
information over all d-entanglements, and the dimensional entropy, which is
achieved at the standard entanglement -- true quantum entanglement,
coinciding with a d-entanglement only in the case of pure marginal states.
The d- and q- information of a quantum noisy channel are respectively
defined via the input d- and q- encodings, and the q-capacity of a quantum
noiseless channel is found as the logarithm of the dimensionality of the
input algebra. The quantum capacity may double the classical capacity,
achieved as the supremum over all d-couplings, or encodings, bounded by the
logarithm of the dimensionality of a maximal Abelian subalgebra.
\end{abstract}

\section{Introduction}

The entanglements, as specifically quantum correlations yet first considered
by Schr\"{o}dinger in \cite{Sch35}, now are used to study quantum
information processes, in particular, quantum computations, quantum
teleportation and quantum cryptography \cite{Ben93,Eke,JS}. There have been
mathematical studies of the entanglements in \cite{Wer89,BBPSSW,Sch1,Wer98},
in which the entangled state of two quantum systems is defined as a compound
state which is not a convex combination $\sum_{n}\varrho _{n}\otimes
\varsigma _{n}p\left( n\right) $ with some states $\varrho _{n}$ and $%
\varsigma _{n}$ on the corresponding algebras $\mathcal{A}$ and $\mathcal{B}$%
. However, it is obvious that there exist several types of correlated
states, written as `separable' forms above. Such correlated, or classically
entangled states have also been discussed in several contexts in quantum
probability, such as quantum measurement and filtering \cite{Bel80, Bel94},
quantum compound states \cite{Ohy83,Ohy83-2} and lifting \cite{AcO}.

In this paper, we study the mathematical structure of classical-quantum and
quantum-quantum couplings to provide a finer classification of quantum
separable and entangled states. We also discuss the informational degree of
entanglement and entangled quantum mutual entropy and quantum capacity. The
latter are treated here solely as quantities arising in certain maximization
problems for quantum mutual information which is generalized here for
arbitrary semifinite algebras.

The term entanglement was introduced by Schr\"{o}dinger in 1935 out of the
need to describe correlations of quantum states not captured by mere
classical statistical correlations which are always the convex combinations
of noncorrelated states. In this spirit the by now standard definition \cite
{Wer89} of the entanglement in physics is the state of a compound quantum
system `which cannot be prepared by two separated devices with only
correlated classical data as their inputs'. We show that the entangled
states can be achieved by quantum (q-) encodings, the nonseparable couplings
of states, in the same way as the separable states can be achieved by
classical (c-) encodings.

The compound states, called o-coupled, are defined by orthogonal
decompositions of their marginal states. This is a particular case of a so
called diagonal (d-compound) state of a compound system which is achieved by
the classical-quantum correspondences called encodings. The d-compound
states as convex combination of the special product states are most
informative among c-compound states, in the sense that maximum of the mutual
entropy over all c-couplings of probe systems $\mathcal{A}$ to the quantum
system $\mathcal{B}$ with a given normal state $\varsigma$ is achieved on
the extreme d-coupled (even o-coupled) states. This maximum is the von
Neumann entropy, which is bound by the rank-capacity $\ln\mathrm{rank}%
\mathcal{B}$, the supremum of $\mathsf{S}\left( \varsigma\right) $ over all $%
\varsigma$. The rank $\mathrm{rank}\mathcal{B}$ of the algebra $\mathcal{B} $
is a topological characteristic of $\mathcal{B}$ defined as the
dimensionality of the maximal Abelian subalgebra $\mathcal{A}\subseteq%
\mathcal{B}$ (in the case of the simple $\mathcal{B}$ it coincides with the
dimensionality $\dim\mathcal{H}$ of the Hilbert space $\mathcal{H}$ of
representation for $\mathcal{B}$). The von Neumann capacity defined as the
maximal von Neumann entropy, i.e. as the maximum $\mathsf{C}_{c}=$ $\ln%
\mathrm{rank}\mathcal{B}$ of mutual entropy over all c-couplings of the
classical probe systems $\mathcal{A}$ to the quantum system $\mathcal{B}$,
is finite only if $\mathrm{rank}\mathcal{B}<\infty$. Due to $\dim\mathcal{B}%
\leq\left( \mathrm{rank}\mathcal{B}\right) ^{2}$ (the equality is only for
the simple algebras $\mathcal{B}$) it is achieved on the normal tracial
density operator $\sigma=\left( \mathrm{rank}\mathcal{B}\right) ^{-1}I$ only
in the case of finite dimensional $\mathcal{B}$.

We prove that the truly entangled compound states are most informative, in
the sense that, the maximum of the mutual entropy over all couplings
including entanglements of the quantum probe systems $\mathcal{A}$ to the
quantum system $\mathcal{B}$ is achieved on a non-separable q-compound
state. It is given by the standard entanglement, an extreme entanglement of $%
\mathcal{A}=\widetilde{\mathcal{B}}$ with the marginal state $\varrho=\tilde{%
\varsigma}$, where $\left( \widetilde{\mathcal{B}},\tilde{\varsigma}\right) $
is the transposed (time inversed) system to $\left( \mathcal{B}%
,\varsigma\right) $. The maximal information gained for such extreme
q-compound states defines another type of entropy, the q-entropy $\mathsf{H}%
\left( \varsigma\right) $, which is bigger than the von Neumann entropy $%
\mathsf{S}\left( \varsigma \right) $ in the case of mixed $\varsigma$. The
maximum of the q-entropy $\mathsf{H}\left( \varsigma\right) $ over all
states $\varsigma$ defines the dimensional capacity $\ln\mathrm{\dim}%
\mathcal{B}$. The dimensionality $\dim\mathcal{B}$ of the algebra $\mathcal{B%
}$ is the major topological characteristic of $\mathcal{B}$, and it gives
true quantum capacity of $\mathcal{B}$ achieved at the standard entanglement
with the maximal chaotic $\varsigma$. Thus, the true quantum capacity is the
maximum $\mathsf{C}_{q}=$ $\ln\dim\mathcal{B}$ of the mutual entropy over
all, not only classical-quantum couplings of the probe systems $\mathcal{A}$
to the quantum system $\mathcal{B}$, and it is finite only for the finite
dimensional algebra $\mathcal{B}$. The q-entropy $\mathsf{H}\left(
\varsigma\right) $, called also the dimensional entropy, can be considered
as the true quantum entropy, in contrast to the von Neumann entropy $\mathsf{%
S}\left( \varsigma\right) $, called also rank-entropy, or c-entropy
(semi-classical entropy) as the supremum of mutual entropy over couplings
only with only classical probe systems $\mathcal{A}$. The capacity $\mathsf{C%
}_{q}$ coincides with $\mathsf{C}_{c}$ only in the classical case of the
Abelian $\mathcal{B}$, and it is strictly larger then the semi-classical
capacity $\mathsf{C}_{c}=\mathrm{\ln rank}\mathcal{B}$ for any noisless
quantum channel. We shall show that the capacity $\mathsf{C}_{q}=\mathrm{%
\ln\dim}\mathcal{B}$ is achieved as the supremum of the quantum Shannon
information for the noisless channel over the entanglements as q-encodings
similar to the capacity $\mathsf{C}_{c}$ which is achieved as the supremum
over c-encodings described by the classical-quantum correspondences $%
\mathcal{A}\rightarrow\mathcal{B}$.

In this paper we consider the case of semifinite quantum systems which are
described by the von Neumann algebras $\mathcal{A}$ and $\mathcal{B}$ with
normal faithful semifinite trace. Such quantum systems include all simple
quantum systems described by full operator algebras as well as all classical
systems as the commutative case. The particular cases of simple and discrete
decomposable algebras are considered in \cite{BeOh98,BeOh00}.

\section{Pairings, Couplings and Entanglements}

In this section we give mathematical characterization of entanglement in
terms of quantum coupling which is described in terms of
transpose-completely positive operations extending individual states to
compound state of a composed quantum system. We show how any normal compound
state can be achieved in this way, and introduce the standard entanglement
as an operation giving rise to the standard entangled compound state.

Let $\mathcal{H}$ denote the Hilbert space of a quantum system, and $%
\mathcal{B}=\mathcal{L}\left( \mathcal{H}\right) $ be the algebra of all
linear bounded operators on $\mathcal{H}$. Note that $\mathcal{B}$ consists
of all operators $A:\mathcal{H}\rightarrow\mathcal{H}$ having the adjoints $%
A^{\dagger}$ on $\mathcal{H}$. A linear functional $\varsigma$ on $\mathcal{B%
}$ with complex-values $\varsigma\left( B\right) \in\mathbb{C}$ is called a
state on $\mathcal{B}$ if it is positive (i.e., $\varsigma\left( B\right)
\geq0$ for any positive operator $B=A^{\dagger}A$ in $\mathcal{B}$) and
normalized (i.e., $\varsigma(I)=1$ for the identity operator $I$ in $%
\mathcal{A}$). A \emph{normal} state can be expressed as, 
\begin{equation}
\varsigma\left( B\right) =\mathrm{Tr}\varkappa^{\dagger}B\varkappa
\equiv\left\langle B,\sigma\right\rangle ,\text{ \quad}B\in\mathcal{B},
\label{1.1}
\end{equation}
where $\varkappa$ is a linear Hilbert-Schmidt operator from $\mathcal{H}$ to
(another) Hilbert space $\mathcal{G}$, and $\varkappa^{\dagger}$ is the
adjoint operator from $\mathcal{G}$ to $\mathcal{H}$. Here $\mathrm{Tr}$
stands for the usual trace in $\mathcal{G}$ (in the case of ambiguity it
will also be denoted as $\mathrm{Tr}_{\mathcal{G}}$). This $\varkappa$ is
called the \emph{amplitude operator} which can always be considered on $%
\mathcal{G}=\mathcal{H}$ as the square root of the operator $%
\varkappa\varkappa^{\dagger }$ (it is called simply the amplitude, if $%
\mathcal{G}$ is the one dimensional space $\mathbb{C}$, $\varkappa=\eta\in%
\mathcal{H}$ with $\varkappa^{\dagger }\varkappa=\Vert\eta\Vert^{2}=1$, in
which case $\varkappa^{\dagger}$ is the functional $\eta^{\dagger}$ from $%
\mathcal{H}$ to $\mathbb{C}$).

We can always equip $\mathcal{H}$ (and will equip all auxiliary Hilbert
spaces, e.g. $\mathcal{G}$) with an isometric involution $J=J^{\dagger}$, $%
J^{2}=I$ having the properties of complex conjugation 
\begin{equation*}
J\sum\lambda_{j}\eta_{j}=\sum\bar{\lambda_{j}}J\eta_{j},\quad\forall
\lambda_{j}\in\mathbb{C},\eta_{j}\in\mathcal{H},
\end{equation*}
and denote by $\left\langle B,\sigma\right\rangle $ the tilde-pairing\ $%
\mathrm{Tr}B\tilde{\sigma}$ of $\mathcal{B}$ with the trace class operators $%
\sigma\in\mathcal{T}\left( \mathcal{H}\right) $ such that $\tilde{\sigma}%
=J\sigma^{\dagger}J$. We shall call $\sigma=J\varkappa \varkappa^{\dagger}J=%
\tilde{\varkappa}^{\dagger}\tilde{\varkappa}$ the probability density of the
state (\ref{1.1}) with respect to this pairing and assume that the support $%
E_{\sigma}$ of $\sigma$ is the minimal projector $E=E^{\dagger}\in\mathcal{B}
$ for which $\varsigma\left( E\right) =1$, i.e. that $\overline{E_{\sigma}}%
:=JE_{\sigma}J=E_{\sigma}$. The latter can also be expressed as the
symmetricity property $\widetilde{E_{\sigma}}=E_{\sigma}$ with respect to
the tilde operation (transposition) $\widetilde{B}=JB^{\dagger}J$ on $%
\mathcal{L}\left( \mathcal{H}\right) $. One can always assume that $J$ is
the standard complex conjugation in an eigen-representation of $\sigma$ such
that $\bar{\sigma}=\varkappa\varkappa^{\dagger}=\tilde {\sigma}$ coincides
with $\sigma$ as the real element of the invariant maximal Abelian
subalgebra $\mathcal{A}\subset\mathcal{L}\left( \mathcal{H}\right) $ of all
diagonal (and thus symmetric) operators in this basis.

The auxiliary Hilbert space $\mathcal{G}$ and the amplitude operator in (\ref
{1.1}) are not unique, however $\varkappa$ is defined uniquely up to a
unitary transform $\varkappa^{\dagger}\mapsto U\varkappa^{\dagger}$ in $%
\mathcal{G}$. $\mathcal{G}$ can always be taken to be minimal by identifying
it with the support $\mathcal{H}_{\sigma}=E_{\sigma}\mathcal{H}$ for $\sigma$
defined as the closure of $\sigma\mathcal{H}$ ($E_{\sigma}$ is the minimal
orthoprojector in $\mathcal{B}$ such that $\sigma E=\sigma)$. In general, $%
\mathcal{G}$ is not one dimensional, the dimensionality $\dim\mathcal{G}$
must not be less than $\mathrm{rank}\varkappa^{\dagger}=\mathrm{rank}\sigma$%
, the dimensionality of the range $\mathrm{ran}\rho=\mathrm{ran}\varkappa
^{\dagger}$ of $\rho=\varkappa^{\dagger}\varkappa$ coinciding with the
support $\mathcal{G}_{\rho}$ for this $\rho\simeq\tilde{\sigma}$.

Given the amplitude operator $\varkappa:\mathcal{G}\rightarrow\mathcal{H}$,
one can define not only the state $\varsigma$ but also the normal state, 
\begin{equation}
\varrho\left( A\right) =\mathrm{Tr}\tilde{\varkappa}^{\dagger}A\tilde{%
\varkappa}\equiv\left\langle A,\rho\right\rangle ,\quad A\in \mathcal{A},
\label{1.2}
\end{equation}
on $\mathcal{A}=\mathcal{L}\left( \mathcal{G}\right) $, as the marginal of
the \emph{pure compound} state

\begin{equation*}
\omega\left( A\otimes B\right) =\mathrm{Tr}\tilde{A}\varkappa^{\dagger
}B\varkappa=\mathrm{Tr}\tilde{\varkappa}^{\dagger}A\tilde{\varkappa}%
\widetilde{B},
\end{equation*}
where $\omega$ is defined on the algebra $\mathcal{A}\otimes\mathcal{B}$ of
all bounded operators on the Hilbert tensor product space $\mathcal{G}\otimes%
\mathcal{H}$.

Indeed, the defined bilinear form, with $\tilde{A}=JA^{\dagger}J$, is
uniquely extended to such a state given on $\mathcal{L}\left( \mathcal{G}%
\otimes\mathcal{H}\right) $ by the amplitude $\psi=\varkappa^{\prime}$,
where $\varkappa^{\prime}$ is uniquely defined by $\left(
\zeta\otimes\eta\right) ^{\dagger}\varkappa^{\prime}=\eta^{\dagger}\varkappa
J\zeta$ for all $\zeta \in\mathcal{G}$, $\eta\in\mathcal{H}$.

This pure compound state $\omega$ is the so called \emph{entangled state} 
\cite{Sch35} unless its marginal state $\varsigma$ (and $\varrho$) is pure
corresponding to a rank one operator $\varkappa^{\dagger}=\zeta\eta^{%
\dagger} $, in which case $\omega=\varrho\otimes\varsigma$ is given by the
amplitude $\upsilon=\zeta\otimes\eta$. The amplitude operator $\varkappa$
corresponding to mixed states on $\mathcal{A}$ and $\mathcal{B}$ will be
called the \emph{entangling operator} of $\rho=\varkappa^{\dagger}\varkappa$
to $\sigma=\tilde{\varkappa}^{\dagger}\tilde{\varkappa}$.

As follows from the next theorem, any pure entangled state 
\begin{equation*}
\omega\left( A\otimes B\right) =\psi^{\dagger}\left( A\otimes B\right)
\psi,\quad A\otimes B\in\mathcal{L}\left( \mathcal{G}\otimes\mathcal{H}%
\right)
\end{equation*}
given by an amplitude $\psi\in\mathcal{G}\otimes\mathcal{H}$, can be
described by a unique entanglement $\varkappa$ to the algebra $\mathcal{A}=%
\mathcal{L}\left( \mathcal{G}\right) $ of the marginal state $\varsigma$ on $%
\mathcal{B}=\mathcal{L}\left( \mathcal{H}\right) $.

Before formulating this theorem in the generality required for further
considerations, let us introduce the following notation.

Let $\mathcal{A}$ be a $\ast$-algebra on $\mathcal{G}$ with a normal,
faithful, semifinite trace $\mu$, $\mathcal{A}^{^{\prime}}$ denote the
commutant $\left\{ A^{\prime}\in\mathcal{L}\left( \mathcal{G}\right) :\left[
A^{\prime},A\right] =0,\forall A\in\mathcal{A}\right\} $ of $\mathcal{A}$,
and $\left( \widetilde{\mathcal{A}},\tilde{\mu}\right) $ denote the
transposed algebra of the operators $\widetilde{A}$ with $\tilde{\mu}\left(
A\right) =\mu\left( \widetilde{A}\right) $, which may not coincide with $%
\left( \mathcal{A},\mu\right) $ (nor with $\mathcal{A}^{\prime}$). We can
always assume that $\widetilde{A}=JA^{\dagger}J$ with respect to an
involution $J$ on $\mathcal{G}$ representing $\widetilde {\mathcal{A}}$ on
the same Hilbert space $\mathcal{G}$ and in most cases $\widetilde{\mathcal{A%
}}=\mathcal{A}$ and $\tilde{\mu}=\mu$ but not in the \emph{standard
representation} unless $\mathcal{A}$ is Abelian algebra. We denote by $%
\mathcal{A}_{\mu}\subseteq\mathcal{A}$ the space of all operators $A\in%
\mathcal{A}$ in the form $x^{\dagger}z$, where $x,z\in\frak{a}_{\mu}$, with $%
\frak{a}_{\mu}=\left\{ x\in\mathcal{A}:\mu\left( x^{\dagger}x\right)
<\infty\right\} $. $\left( \mathcal{G}_{\mu},\iota,J_{\mu}\right) $ denotes
the standard representation $\iota:\mathcal{A}\rightarrow\mathcal{L}\left( 
\mathcal{G}_{\mu}\right) $ given by the left multiplication $\iota\left(
A\right) x=Ax$ on $\frak{a}_{\mu}$, with the standard isometric involution $%
J_{\mu}:x\mapsto x^{\dagger}$ defining the representation $\tilde{\iota }%
\left( \widetilde{A}\right) =J_{\mu}\iota\left( A^{\dagger}\right) J_{\mu}$
of $\widetilde{\mathcal{A}}$ on the completion $\mathcal{G}_{\mu}$ of the
module $\frak{a}_{\mu}$ with respect to the inner product $\left( x|z\right)
_{\mu}=\mu\left( x^{\dagger}z\right) $. We recall that the von Neumann
algebra $\mathcal{A}$ defined by $\mathcal{A}^{\prime\prime }=\mathcal{A}$
is anti-isomorphic to $\iota\left( \mathcal{A}\right)
^{\prime}=J_{\mu}\iota\left( \mathcal{A}\right) J_{\mu}$ and thus $%
\widetilde{\mathcal{A}}\simeq\iota\left( \mathcal{A}\right) ^{\prime}$ and
that $\widetilde{\mathcal{A}}=\mathcal{A}_{\mu}^{\ast}$ as the space of all
continuous functionals $\widetilde{A}:\phi\mapsto\left\langle \phi ,%
\widetilde{A}\right\rangle _{\mu}$ with respect to the $\ast$-norm $\left\|
\phi\right\| _{\ast}=\sup\left\{ \left| \mu\left( A\phi\right) \right|
:\left\| A\right\| \leq1\right\} $ on $\mathcal{A}_{\mu}$ and the pairing 
\begin{equation*}
\left\langle x^{\dagger}z,\widetilde{A}\right\rangle _{\mu}=\mu\left(
zAx^{\dagger}\right) =\left\langle A,\widetilde{x^{\dagger}z}\right\rangle
_{\mu},\quad x^{\dagger}z\in\mathcal{A}_{\mu},.\widetilde{A}\in\widetilde {%
\mathcal{A}}.
\end{equation*}
The completion of $\mathcal{A}_{\mu}$ with respect to the norm $\left\|
\cdot\right\| _{\ast}$ is the predual Banach space, denoted as $\mathcal{A}%
_{\ast}$ ( if $\mu=\tau|\mathcal{A}$ is the usual trace $\tau=\mathrm{Tr}_{%
\mathcal{G}}$ on $\mathcal{A}$, then $\mathcal{A}_{\mu}$ coincides with $%
\mathcal{A}_{\ast}$ as the class $\mathcal{A}_{\tau}=\mathcal{A}\cap\mathcal{%
T}\left( \mathcal{G}\right) $ of trace operators $\mathcal{T}\left( \mathcal{%
G}\right) =\left\{ x^{\dagger}z:x,z\in\mathcal{S}\left( \mathcal{G}\right)
\right\} $, where $\mathcal{S}\left( \mathcal{G}\right) =\left\{ x\in%
\mathcal{L}\left( \mathcal{G}\right) :\mathrm{Tr}_{\mathcal{G}%
}x^{\dagger}x<\infty\right\} $).

If $\mathcal{A}$ is not the algebra of all operators $\mathcal{L}\left( 
\mathcal{G}\right) $, the density operator $\rho$ for a normal state (\ref
{1.2}) is not unique with respect to $\tau=\mathrm{Tr}_{\mathcal{G}}$.
However it is uniquely defined as the bounded \emph{probability density} $%
\rho=Jx^{\dagger}xJ=\bar{x}^{\dagger}\bar{x}$ with respect to the
restriction $\mu=\tau|\mathcal{A}$ (i.e. as the density operator with
respect to $\mu$) describing this state as $\left\langle A,\rho\right\rangle
_{\mu}=\mu\left( xAx^{\dagger}\right) $ by the additional condition $%
\varkappa=\bar{x}\in\widetilde{\mathcal{A}_{\mu}}$. Note that each
probability density $\rho \in\widetilde{\mathcal{A}_{\mu}}$ describing the
normal state $\varrho\left( A\right) =\left\langle A,\rho\right\rangle
_{\mu} $ on $\mathcal{A}\ni A$ is positive and normalized as $\left\langle
I,\rho\right\rangle _{\mu}=1$. However the predual space $\widetilde{%
\mathcal{A}}_{\ast}$ as the $\ast $-completion of $\widetilde{\mathcal{A}%
_{\mu}}$ may consist of not only the bounded densities with respect to $\mu$
(however each $\rho\in\widetilde {\mathcal{A}}_{\ast}$ can always be
approximated by the bounded $\rho_{n}\in\widetilde{\mathcal{A}_{\mu}}$).

In the following formulation $\mathcal{B}$ can also be the more general von
Neumann algebra, rather than $\mathcal{L}\left( \mathcal{H}\right) $, with a
normal faithful semifinite trace $\nu :\mathcal{B}_{\nu }\mapsto \mathbb{C}$
defining the pairing $\left\langle B,u^{\dagger }u\right\rangle _{\nu }=\nu
\left( \tilde{u}^{\dagger }B\tilde{u}\right) $, where $u\in \widetilde{\frak{%
b}_{\nu }}$ ($\mathcal{B}_{\nu }=\frak{b}_{\nu }^{\dagger }\frak{b}_{\nu }$
coincides with $\mathcal{B}_{\ast }$ in the case of the standard trace $\nu
\left( B\tilde{\sigma}\right) =\mathrm{Tr}B\tilde{\sigma}=\left\langle
B,\sigma \right\rangle _{\nu }$ when $\frak{b}_{\nu }$ is the space of
Hilbert-Schmidt operators $y\in \mathcal{B}$ and $\widetilde{\mathcal{B}}=%
\mathcal{B}$).

\begin{theorem}
Let $\omega :\mathcal{A}\otimes \mathcal{B}\rightarrow \mathbb{C}$ be a
normal compound state 
\begin{equation}
\omega \left( A\otimes B\right) =\tau \left( \upsilon \widetilde{\left(
A\otimes B\right) }\upsilon ^{\dagger }\right) :=\left\langle A\otimes
B,\upsilon ^{\dagger }\upsilon \right\rangle ,  \label{1.3}
\end{equation}
described by an amplitude operator $\upsilon :\mathcal{G}\otimes \mathcal{H}%
\rightarrow \mathcal{E}\otimes \mathcal{F}$ on the tensor product of Hilbert
spaces $\mathcal{E}$ and $\mathcal{F}$, satisfying the condition 
\begin{equation*}
\upsilon ^{\dagger }\upsilon \in \widetilde{\mathcal{A}}\otimes \widetilde{%
\mathcal{B}},\quad \tau \left( \upsilon \upsilon ^{\dagger }\right) =1,
\end{equation*}
where $\tau \simeq \tilde{\mu}\otimes \tilde{\nu}$ is the trace $\tau \left(
\upsilon \upsilon ^{\dagger }\right) =\left\langle I\otimes I,\upsilon
^{\dagger }\upsilon \right\rangle $ defined in (\ref{1.3}) by the pairing
for $\mathcal{A}\otimes \mathcal{B}$ with respect to $\mu \otimes \nu $.
Then this state is achieved by an entangling operator $\varkappa :\mathcal{G}%
\otimes \mathcal{F}\rightarrow \mathcal{E}\otimes \mathcal{H}$ as 
\begin{equation}
\left\langle A,\nu \left( \varkappa ^{\dagger }\left( I\otimes B\right)
\varkappa \right) \right\rangle _{\mu }=\omega \left( A\otimes B\right)
=\left\langle B,\mu \left( \tilde{\varkappa}^{\dagger }\left( A\otimes
I\right) \tilde{\varkappa}\right) \right\rangle _{\nu }  \label{1.4}
\end{equation}
for all $A\in \mathcal{A}$ and $B\in \mathcal{B}$ such that 
\begin{equation*}
\nu \left( \varkappa ^{\dagger }\left( I\otimes B\right) \varkappa \right)
\subseteq \widetilde{\mathcal{A}},\quad \mu \left( \tilde{\varkappa}%
^{\dagger }\left( A\otimes I\right) \tilde{\varkappa}\right) \subseteq 
\widetilde{\mathcal{B}}.
\end{equation*}
The operator $\varkappa $ together with $\tilde{\varkappa}=J\varkappa
^{\dagger }J$ is uniquely defined by $\upsilon =U\varkappa ^{\prime }$,
where 
\begin{equation}
\left( \xi \otimes \eta ^{\prime }\right) ^{\dagger }\varkappa ^{\prime
}\left( \zeta \otimes J\eta \right) =\left( \xi \otimes \eta \right)
^{\dagger }\varkappa \left( \zeta \otimes J\eta ^{\prime }\right) ,\quad \xi
\in \mathcal{E},\eta ^{\prime }\in \mathcal{F},\zeta \in \mathcal{G},\eta
\in \mathcal{H},  \label{1.7}
\end{equation}
up to a unitary transformation $U$ of the minimal subspace space $\mathrm{ran%
}\upsilon \subseteq \mathcal{E}\otimes \mathcal{F}$.
\end{theorem}

\proof%
%
Without loss of generality we can assume that $\mathcal{E}=\mathcal{G}_{\rho
}$, $\mathcal{F}=\mathcal{H}_{\sigma }$ and $\upsilon ^{\dagger }=\upsilon
\left( E_{\rho }\otimes E_{\sigma }\right) $ as the support $\left( \mathcal{%
G}\otimes \mathcal{H}\right) _{\upsilon ^{\dagger }\upsilon }=\mathrm{ran}%
\upsilon ^{\dagger }$ for $\upsilon ^{\dagger }\upsilon $ is contained in $%
\mathcal{G}_{\rho }\otimes \mathcal{H}_{\sigma }$. Due to $\upsilon
^{\dagger }\upsilon \in \left( \widetilde{\mathcal{A}^{\prime }}\otimes 
\widetilde{\mathcal{B}^{\prime }}\right) ^{\prime }$ the range of $\upsilon $
is invariant under the action 
\begin{equation*}
\left( A\otimes B\right) \upsilon =\upsilon \left( AE_{\rho }\otimes
BE_{\sigma }\right) ,\quad \forall A\in \widetilde{\mathcal{A}^{\prime }}%
,B\in \widetilde{\mathcal{B}^{\prime }}
\end{equation*}
of the commutant $\left( \widetilde{\mathcal{A}}\otimes \widetilde{\mathcal{B%
}}\right) ^{\prime }=\widetilde{\mathcal{A}^{\prime }}\otimes \widetilde{%
\mathcal{B}^{\prime }}$. Let us equip $\mathcal{G}$ and $\mathcal{H}$ with
the involutions $J$ leaving invariant $\mathcal{G}_{\rho }=E_{\rho }\mathcal{%
G}$ and $\mathcal{H}_{\sigma }=E_{\sigma }\mathcal{H}$ denoting $J_{\rho
}=E_{\rho }J$, $J_{\sigma }=E_{\sigma }J$, and $\mathcal{E}\otimes \mathcal{F%
}=\mathcal{G}_{\rho }\otimes \mathcal{H}_{\sigma }$ with the induced
involution $J\left( \zeta \otimes \eta \right) =J_{\rho }\zeta \otimes
J_{\sigma }\eta $. It is easy to check for such $\upsilon $ and $\varkappa
=\upsilon ^{\prime }$ defined by $\upsilon =\varkappa ^{\prime }$ in (\ref
{1.7}) that for any $A\in \mathcal{A}^{\prime }$ and $B\in \widetilde{%
\mathcal{B}^{\prime }}$%
\begin{align*}
\left( \widetilde{A}\xi \otimes \eta \right) ^{\dagger }\varkappa \left(
\zeta \otimes \overline{B}J\eta ^{\prime }\right) & =\left( \widetilde{A}\xi
\otimes B\eta ^{\prime }\right) ^{\dagger }\upsilon \left( \xi \otimes J\eta
\right) =\left( \xi \otimes \eta ^{\prime }\right) ^{\dagger }\upsilon
\left( \overline{A}\xi \otimes J\widetilde{B}\eta \right) \\
& =\left( \xi \otimes \widetilde{B}\eta \right) ^{\dagger }\varkappa \left( 
\overline{A}\zeta \otimes J\eta ^{\prime }\right)
\end{align*}
where $\overline{A}=JAJ\in \widetilde{\mathcal{A}^{\prime }}$, $\overline{B}%
=JBJ\in \mathcal{B}^{\prime }$. Hence for any $B\in \mathcal{B}$ 
\begin{equation*}
\left( A\otimes B^{\prime }\right) \varkappa ^{\dagger }\left( I\otimes
B\right) \varkappa =\varkappa ^{\dagger }\left( A\otimes B^{\prime }B\right)
\varkappa =\varkappa ^{\dagger }\left( I\otimes B\right) \varkappa \left(
A\otimes B^{\prime }\right) ,
\end{equation*}
where $A\in \widetilde{\mathcal{A}_{\rho }^{\prime }}:=\widetilde{\mathcal{A}%
^{\prime }}E_{\rho }$, $B^{\prime }\in \mathcal{B}_{\sigma }^{\prime }:=%
\mathcal{B}^{\prime }E_{\sigma }$, and for any $A\in \mathcal{A}$%
\begin{equation*}
\left( A^{\prime }\otimes B\right) \tilde{\varkappa}^{\dagger }\left(
A\otimes I\right) \tilde{\varkappa}=\varkappa \left( A^{\prime }A\otimes
B\right) \varkappa ^{\dagger }=\tilde{\varkappa}^{\dagger }\left( A\otimes
I\right) \tilde{\varkappa}\left( A^{\prime }\otimes B\right) ,
\end{equation*}
where $A^{\prime }\in \mathcal{A}^{\prime }$ and $B\in \widetilde{\mathcal{B}%
^{\prime }}$. Thus for all $A\in \mathcal{A}$ and $B\in \mathcal{B}$ 
\begin{equation*}
\varkappa ^{\dagger }\left( I\otimes B\right) \varkappa \in \left( 
\widetilde{\mathcal{A}_{\rho }^{\prime }}\otimes \mathcal{B}_{\sigma
}^{\prime }\right) ^{\prime },\quad \tilde{\varkappa}^{\dagger }\left(
A\otimes I\right) \tilde{\varkappa}\in \left( \mathcal{A}_{\rho }^{\prime
}\otimes \widetilde{\mathcal{B}_{\sigma }^{\prime }}\right) ^{\prime }.
\end{equation*}
Moreover, due to $\mathcal{A}_{\rho }^{\prime \prime }=E_{\rho }\mathcal{A}%
E_{\rho }\equiv \mathcal{A}_{\rho }$ and $\mathcal{B}_{\sigma }^{\prime
\prime }=E_{\sigma }\mathcal{B}E_{\sigma }\equiv \mathcal{B}_{\sigma }$%
\begin{equation*}
\varkappa ^{\dagger }\left( I\otimes B\right) \varkappa \subseteq J_{\rho }%
\mathcal{A}_{\mu }J_{\rho }\otimes E_{\sigma }\mathcal{B}_{\nu }E_{\sigma
}:=\left( \widetilde{\mathcal{A}_{\rho }}\otimes \mathcal{B}_{\sigma
}\right) _{\tilde{\mu}\otimes \nu },
\end{equation*}
\begin{equation*}
\tilde{\varkappa}^{\dagger }\left( A\otimes I\right) \tilde{\varkappa}%
\subseteq E_{\rho }\mathcal{A}_{\mu }E_{\rho }\otimes J_{\sigma }\mathcal{B}%
_{\nu }J_{\sigma }:=\left( \mathcal{A}_{\rho }\otimes \widetilde{\mathcal{B}%
_{\sigma }}\right) _{\mu \otimes \tilde{\nu}}
\end{equation*}
as bounded by $\left\| B\right\| \varkappa ^{\dagger }\varkappa $ and by $%
\left\| A\right\| \tilde{\varkappa}^{\dagger }\tilde{\varkappa}$
respectively. The partial traces $\nu $ and $\mu $ on these reduced algebras
are defined as 
\begin{equation}
\nu \left( \varkappa ^{\dagger }\left( I\otimes B\right) \varkappa \right)
=\left\langle B,\upsilon ^{\dagger }\upsilon \right\rangle _{\nu },\;\mu
\left( \tilde{\varkappa}^{\dagger }\left( A\otimes I\right) \tilde{\varkappa}%
\right) =\left\langle A,\upsilon ^{\dagger }\upsilon \right\rangle _{\mu },
\label{1.5}
\end{equation}
according to $\left\langle A,\left\langle B,\upsilon ^{\dagger }\upsilon
\right\rangle _{\nu }\right\rangle _{\mu }=\left\langle A\otimes B,\upsilon
^{\dagger }\upsilon \right\rangle =\left\langle B,\left\langle A,\upsilon
^{\dagger }\upsilon \right\rangle _{\mu }\right\rangle _{\nu }$, where 
\begin{equation*}
\left\langle B,\upsilon ^{\dagger }\upsilon \right\rangle _{\nu }=\widetilde{%
\nu \left( \left( I\otimes B\right) \widetilde{\upsilon ^{\dagger }\upsilon }%
\right) },\quad \left\langle A,\upsilon ^{\dagger }\upsilon \right\rangle
_{\mu }\widetilde{=\mu \left( \left( A\otimes I\right) \widetilde{\upsilon
^{\dagger }\upsilon }\right) }
\end{equation*}
In particular 
\begin{equation*}
\nu \left( \varkappa ^{\dagger }\varkappa \right) =\tilde{\nu}\left(
\upsilon ^{\dagger }\upsilon \right) =\rho ,\;\mu \left( \tilde{\varkappa}%
^{\dagger }\tilde{\varkappa}\right) =\tilde{\mu}\left( \upsilon ^{\dagger
}\upsilon \right) =\sigma .
\end{equation*}
Any other choice of $\upsilon $ with the minimal $\mathcal{E}\otimes 
\mathcal{F}\simeq \mathcal{G}_{\rho }\otimes \mathcal{H}_{\sigma }$ is
unitary equivalent to $\varkappa ^{\prime }$.%
\endproof%
%

Note that the entangled state (\ref{1.3}) is written in (\ref{1.4}) as, 
\begin{equation*}
\left\langle B,\varpi\left( A\right) \right\rangle _{\nu}=\omega\left(
A\otimes B\right) =\left\langle A,\varpi^{\intercal}\left( B\right)
\right\rangle _{\mu},
\end{equation*}
in terms of the mutually adjoint maps $\varpi:\mathcal{A}\rightarrow 
\widetilde{\mathcal{B}}_{\ast}$ and $\varpi^{\intercal}:\mathcal{B}%
\rightarrow\widetilde{\mathcal{A}}_{\ast}$. These maps are given in (\ref
{1.5}) as 
\begin{equation}
\varpi\left( A\right) =\left\langle A,\upsilon^{\dagger}\upsilon
\right\rangle _{\mu}=\widetilde{\pi^{\ast}\left( A\right) },\quad
\varpi^{\intercal}\left( B\right) =\left\langle
B,\upsilon^{\dagger}\upsilon\right\rangle _{\nu}=\widetilde{\pi\left(
B\right) },  \label{1.6}
\end{equation}
where the linear map $\pi:\mathcal{B}\rightarrow\mathcal{A}_{\mu}$ and the
adjoint $\pi^{\ast}:\mathcal{A}\rightarrow\mathcal{B}_{\nu}$ are defined as
partial traces 
\begin{equation*}
\pi\left( B\right) =\nu\left( \left( I\otimes B\right) \widetilde
{\upsilon^{\dagger}\upsilon}\right) ,\quad\pi^{\ast}\left( A\right)
=\mu\left( \left( A\otimes I\right) \widetilde{\upsilon^{\dagger}\upsilon }%
\right) .
\end{equation*}

The linear normal map $\varpi$ in (\ref{1.5}) is written in the
Kraus-Stinespring form \cite{Sti55} and thus is completely positive (CP). It
is not unital but normalized to the density operators $\sigma=\omega\left(
I\right) $ with respect to the weight $\nu$.

A linear map $\pi:\mathcal{B}\rightarrow\mathcal{A}_{\ast}$ is called \emph{%
tilde-positive} if the map $\pi^{\symbol{126}}$ defined as $\pi^{\symbol{126}%
}\left( B\right) :=J\pi\left( B\right) ^{\dagger}J$ is positive for any
positive (and thus Hermitian) operator $B\geq0$ in the sense of non-negative
definiteness of $B$. It is called \emph{tilde-completely positive} (TCP) if
the operator-matrix $\pi^{\symbol{126}}\left( \mathbf{B}\right) =J\pi\left( 
\mathbf{B}\right) ^{\dagger}J$ is positive for every positive
operator-matrix $\mathbf{B}=\left[ B_{ik}\right] =\mathbf{B}^{\ast}$, where $%
\mathbf{A}^{\dagger}=\left[ A_{ik}^{\dagger }\right] $, $\mathbf{B}^{\ast}=%
\left[ B_{ki}^{\dagger}\right] $ (and thus $\mathbf{A}^{\dagger}=\left[
A_{ki}\right] $ for $\mathbf{A=}\left[ A_{ik}\right] \geq0$, and $\mathbf{B}%
^{\ast}=\mathbf{B}$ for $\mathbf{B}\geq0$). Obviously every tilde-positive
and tilde-completely positive $\pi$ is positive as positive is $\tilde{A}%
=JA^{\dagger}J$ for every positive $A$, but it is not necessarily completely
positive unless $\tilde{A}=A$ for all $A\in\mathcal{A}$, in which case $%
\mathcal{A}$ is Abelian (or the Abelian is $\mathcal{B}$).

The map $\pi$ defined in (\ref{1.8}) as a TCP $\dagger$-map, $\pi\left(
B^{\dagger}\right) =\pi\left( B\right) ^{\dagger}$, is obviously
transpose-CP in the sense of positivity of $\pi\left( \mathbf{B}\right)
^{\dagger}=\left[ \pi\left( B_{ki}\right) \right] =\pi\left( \mathbf{B}%
^{\dagger}\right) $ for any $\mathbf{B}\geq0$, but it is in general not CP$.$
Because every transpose-CP map can be represented as tilde-CP: there might
be a positive-definite matrix $\mathbf{B}$ for which $\pi\left( \mathbf{B}%
\right) $ is not positive. Note that the adjoint map $\pi^{\ast }=\bar{\pi}%
^{\intercal}$ is also TCP, as well as the maps $\tilde{\pi}=\bar{\pi}$ and $%
\pi^{\intercal}=\bar{\pi}^{\ast}$, where $\bar{\pi}\left( B\right)
=J\pi\left( \overline{B}\right) J$, obtained from (\ref{1.5}) as partial
tracings 
\begin{equation}
\bar{\pi}\left( B\right) =\nu\left( \varkappa^{\dagger}\left( I\otimes%
\widetilde{B}\right) \varkappa\right) ,\quad\pi^{\intercal}\left( A\right)
=\mu\left( \tilde{\varkappa}^{\dagger}\left( \widetilde{A}\otimes I\right) 
\tilde{\varkappa}\right) .  \label{1.8}
\end{equation}
In these terms, the compound state (\ref{1.4}) is written as, 
\begin{equation*}
\left\langle A|\pi\left( B\right) \right\rangle _{\mu}=\omega\left(
A^{\dagger}\otimes B\right) =\left\langle \pi^{\ast}\left( A\right)
|B\right\rangle _{\nu},
\end{equation*}
where $\left\langle x|y\right\rangle =\left\langle y,\overline{x}%
\right\rangle $ defines an inner product which coincides in the case of
traces with the GNS product $\left( x|y\right) $.

In the following definition the predual space $\mathcal{B}_{\intercal }=%
\widetilde{\mathcal{B}}_{\ast }$ (as well as $\mathcal{A}_{\intercal }=%
\widetilde{\mathcal{A}}_{\ast }$) is identified by the pairing $\left\langle
B,\sigma \right\rangle _{\nu }=\varsigma \left( B\right) $ with the space of
generalized density operators $\sigma $ which are thus uniquely defined as
selfadjoint, in general unbounded, operators in $\mathcal{H}$. Note that $%
\mathcal{B}_{\intercal }=\mathcal{B}_{\nu }$ if $\mathcal{B}=\widetilde{%
\mathcal{B}}$ and $\nu =\mathrm{Tr}_{\mathcal{H}}=\tilde{\nu}$.

\begin{definition}
A TCP map $\pi :\mathcal{B}\rightarrow \mathcal{A}_{\ast }$ (or $\mathcal{B}%
\rightarrow \mathcal{A}_{\mu }\subseteq \mathcal{A}_{\ast }$) normalized as $%
\mu \left( \pi \left( I\right) \right) =1$ and having an adjoint with $\pi
^{\ast }\left( \mathcal{A}\right) \subseteq \mathcal{B}_{\ast }$ ($\pi
^{\ast }\left( \mathcal{A}\right) \subseteq \mathcal{B}_{\nu }$) is called
normal coupling (bounded coupling) of the state $\varsigma =\mu \circ \pi $
on $\mathcal{B}$ to the state $\varrho =\nu \circ \pi ^{\ast }$ on $\mathcal{%
A}$. The CP map $\varpi :\mathcal{A}\rightarrow \mathcal{B}_{\intercal }$
(or $\mathcal{A}\rightarrow \widetilde{\mathcal{B}_{\nu }}\subseteq \mathcal{%
B}_{\intercal }$) normalized to the probability\ density $\sigma =\varpi
\left( I\right) $ of $\varsigma $ with $\varpi ^{\intercal }\left( I\right)
\in \mathcal{B}_{\ast }$ ($\varpi ^{\intercal }\left( I\right) \in 
\widetilde{\mathcal{A}_{\mu }}$) will be called normal entanglement (bounded
entanglement) of the system $\left( \mathcal{A},\varrho \right) $ with the
probability density $\rho =\varpi ^{\intercal }\left( I\right) $ to $\left( 
\mathcal{B},\varsigma \right) $. The coupling $\pi $ (entanglement $\varpi $%
) is called truly quantum if it is not CP (not TCP). The self-adjoint
entanglement $\varpi _{q}=\varpi _{q}^{\ast }$ on $\left( \mathcal{A}%
,\varrho \right) =\left( \widetilde{\mathcal{B}},\tilde{\varsigma}\right) $
(or symmetric coupling $\pi _{q}=\pi _{q}^{\intercal }$ into $\mathcal{A}%
_{\ast }=\mathcal{B}_{\intercal }$)$\ $is called standard for the system $%
\left( \mathcal{B},\varsigma \right) $ if it is given by 
\begin{equation}
\varpi _{q}\left( A\right) =\sigma ^{1/2}A\sigma ^{1/2},\quad \pi _{q}\left(
B\right) =\sigma ^{1/2}\widetilde{B}\sigma ^{1/2}.  \label{1.9}
\end{equation}
\end{definition}

Note that the standard entanglement is true as soon as the reduced algebra $%
\mathcal{B}_{\sigma}=E_{\sigma}\mathcal{B}E_{\sigma}$ on the support $%
\mathcal{H}_{\sigma}=E_{\sigma}\mathcal{H}$ of the state $\varsigma$ is not
Abelian, i.e. is not one-dimensional in the case $\mathcal{B}=\mathcal{L}%
\left( \mathcal{H}\right) $, corresponding to a pure normal $\varsigma$ on $%
\mathcal{B=L}\left( \mathcal{H}\right) $. Indeed, $\pi^{q}$ restricted to $%
\mathcal{B}_{\sigma}$ is the composition of the nondegenerated
multiplication $\mathcal{B}_{\sigma}\ni B\mapsto\tilde{\sigma}^{1/2}B$ $%
\tilde{\sigma}^{1/2}$ (which is CP) and the transposition $\widetilde{B}%
=JB^{\dagger}J$ on $\mathcal{B}_{\sigma}$ (which is TCP but not CP if $\dim%
\mathcal{H}_{\sigma }>1$).

The standard entanglement in the purely quantum case $\mathcal{B=B}\left( 
\mathcal{H}\right) =\widetilde{\mathcal{B}}$, $\nu=\mathrm{Tr}=\tilde{\nu}$
corresponds to the pure \emph{standard compound state} 
\begin{equation}
\mathrm{Tr}A\sigma^{1/2}\widetilde{B}\sigma^{1/2}=\omega_{q}\left( A\otimes
B\right) =\mathrm{Tr}B\tilde{\sigma}^{1/2}\widetilde{A}\tilde{\sigma}^{1/2}
\label{1.10}
\end{equation}
on the algebra $\mathcal{B}\otimes\mathcal{B}$. It is given by the amplitude 
$\upsilon^{\prime}\simeq|\sigma^{1/2})\equiv\psi$, with $|\sigma
^{1/2})^{\dagger}=\varkappa^{\prime}\equiv(\sigma^{1/2}|$ defined in (\ref
{1.7}) as $\varkappa^{\prime}\left( \zeta\otimes J\eta\right)
=\eta^{\dagger}\varkappa\zeta$ for $\varkappa=\sigma^{1/2}$.

Any entanglement on $\mathcal{A}=\mathcal{L}\left( \mathcal{G}\right) $, $%
\mu=\mathrm{Tr}$ corresponding to a pure compound state is true if $\mathrm{%
rank}\rho=\mathrm{rank}\sigma$ is not one. If the space $\mathcal{G}$ is
also minimal, $\mathcal{G}=\mathcal{G}_{\rho}$, $\pi^{\intercal}$ is unitary
equivalent to the standard one $\pi_{q}$. Indeed, $\varpi\left( A\right) =%
\tilde{\varkappa}^{\dagger}A\tilde{\varkappa}$ can be decomposed as 
\begin{equation*}
\varpi\left( A\right) =\sigma^{1/2}U^{\dagger}AU\sigma^{1/2}=\varpi
_{q}\left( U^{\dagger}AU\right) ,
\end{equation*}
where $U:\sigma^{1/2}\eta\mapsto\tilde{\varkappa}\eta$ is a unitary operator
from $\mathcal{H}_{\sigma}$ onto the support $\mathcal{G}_{\rho}$ of $%
\rho=U\sigma U^{\dagger}$ with nonabelian $\mathcal{A}_{\rho}=\mathcal{L}%
\left( \mathcal{G}_{\rho}\right) $ and $\mathcal{B}_{\sigma}=U^{\dagger }%
\mathcal{A}_{\rho}U=\mathcal{L}\left( \mathcal{H}_{\sigma}\right) $.

Note that the compound state (\ref{1.4}) with $\tilde{\varkappa}%
=\sigma^{1/2} $ corresponding to the standard $\varpi=\varpi_{q}$ can always
be extended to a vector state on $\widetilde{\mathcal{B}}\vee\mathcal{B}$ in
the standard representation $\left( \mathcal{H}_{\nu},\iota,J_{\nu}\right) $%
\ of $\mathcal{B}\equiv\iota\left( \mathcal{B}\right) $ when $\widetilde {%
\mathcal{B}}=J_{\nu}\mathcal{B}J_{\nu}=\mathcal{B}^{\prime}$. However it
cannot be extended to a normal state on $\widetilde{\mathcal{B}}\otimes%
\mathcal{B}$ in the case of nonatomic $\mathcal{B}$. If $\mathcal{B}$ is a
factor this state is pure, given in the standard representation $\widetilde{%
\mathcal{B}}\vee\mathcal{B}=\mathcal{L}\left( \mathcal{H}_{\nu }\right) $ by
the unit vector $y=\tilde{\sigma}^{1/2}\in\mathcal{H}_{\nu}$; however it is
not normal on $\widetilde{\mathcal{B}}\otimes\mathcal{B}$ unless $\mathcal{B}
$ is type I: $\mathcal{B}\simeq\mathcal{L}\left( \mathcal{H}\right) $.

\section{C-, D- and O-Couplings and Encodings}

In this section we discuss the operational meaning of couplings
corresponding to different types of encodings which are treated here solely
in terms of coupling maps on input of a quantum physical system. We hope
that this mathematical treatment will provide a new physical insight for the
corresponding asymptotic problems of quantum information.

The compound states play the role of joint input-output probability measures
in classical information channels and can be pure in the quantum case, even
if the marginal states are mixed. The pure compound states achieved by an
entanglement of mixed input and output states exhibit new, non-classical
type correlations, which are responsible for the EPR type paradoxes in the
interpretation of quantum theory \cite{Wer89}. However, mixed, so called 
\emph{separable} states on $\mathcal{A}\otimes\mathcal{B}$, defined as
convex product combinations 
\begin{equation*}
\omega_{c}\left( A\otimes B\right) =\sum_{n}\varrho_{n}\left( A\right)
\varsigma_{n}\left( B\right) p\left( n\right) ,
\end{equation*}
which we refer as the \emph{c-compound states}, do not exhibit such
paradoxical behavior. Here $p\left( n\right) >0$, $\sum p_{n}=1$, is a
probability distribution, and $\varrho_{n}:\mathcal{A}\rightarrow\mathbb{C} $%
, $\varsigma_{n}:\mathcal{B}\rightarrow\mathbb{C}$ are usually normal states
defined by the product densities $\rho_{n}\otimes\sigma_{n}\in$ $\mathcal{A}%
_{\intercal}\otimes\mathcal{B}_{\intercal}$ of $\omega_{n}=\varrho_{n}%
\otimes\sigma_{n}$. Such compound states are achieved by \emph{c-couplings} $%
\pi_{c}:\mathcal{B}\rightarrow\mathcal{A}_{\ast}$ given by $\pi _{c}^{%
\symbol{126}}=\varpi_{c}^{\intercal}$, where 
\begin{equation*}
\varpi_{c}\left( A\right) =\sum_{n}\varrho_{n}\left( A\right) \sigma
_{n}p\left( n\right) ,\;\varpi_{c}^{\intercal}\left( B\right) =\sum
_{n}\varsigma_{n}\left( B\right) \rho_{n}p\left( n\right) ,
\end{equation*}
Here $\rho_{n}\in\mathcal{A}_{\ast}$ and $\sigma_{n}\in\mathcal{B}_{\ast}$
are the probability densities for $\varrho_{n}$ and $\varsigma_{n}$ with
respect to given traces $\mu$ and $\nu$ on $\mathcal{A}$ and $\mathcal{B}$.
Note that the \emph{c-entanglement} $\varpi_{c}$, being the convex
combinations of the primitive CP-TCP\ maps $\varpi_{n}\left( A\right)
=\varrho_{n}\left( A\right) \sigma_{n}\in\mathcal{B}_{\intercal} $, is not
truly quantum.

The separable states of the particular form 
\begin{equation}
\omega_{d}\left( A\otimes B\right) =\sum_{n}\langle n|A|n\rangle
\varsigma\left( n,B\right) ,  \label{2.1}
\end{equation}
where $\varrho_{n}\left( A\right) =\langle n|A|n\rangle$ are pure states on $%
\mathcal{A}=\mathcal{L}\left( \mathcal{G}\right) =\widetilde{\mathcal{A}} $
given by an ortho-normal system $\left\{ |n\rangle\right\} \subset \mathcal{G%
}$, and $\varsigma\left( n,B\right) =\left\langle B,\sigma\left( n\right)
\right\rangle _{\nu}$ with $\sigma\left( n\right) =\sigma _{n}p\left(
n\right) $, are usually considered as the proper candidates for the
input-output states in the communication channels involving the
classical-quantum (c-q) encodings. Such a separable state was introduced by
Ohya \cite{Ohy83,Ohy89} using a Schatten decomposition $\rho=\sum
|n\rangle\langle n|p\left( n\right) $ of the input density operator $\rho \in%
\mathcal{T}\left( \mathcal{G}\right) $ into the orthogonal one-dimensional
projectors $\rho_{n}=|n\rangle\langle n|$. Here we note that such a state is
the mixture of the classical-quantum correspondences $n\mapsto|n\rangle%
\langle n|\otimes\sigma_{n}$ which can be described as the composition of
quantum channeling $|n\rangle\langle n|\mapsto\sigma_{n}$ and the errorless 
\emph{encodings} $n\mapsto|n\rangle\langle n|$ in the sense that they can be
inverted by the measurements $|n\rangle\langle n|\mapsto n$ as input \emph{%
decodings}. We shall call such separable states \emph{d-compound} as they
are achieved by the diagonal couplings $\pi_{d}=\varpi_{d}^{\intercal }$ (%
\emph{d-couplings}) to the subalgebra $\mathcal{A}_{d}\subseteq \mathcal{A}$
of the diagonal operators $A=\sum a\left( n\right) |n\rangle\langle n|$,
where 
\begin{equation}
\varpi_{d}\left( A\right) =\sum_{n}\langle n|A|n\rangle\sigma\left( n\right)
,\;\varpi_{d}^{\intercal}\left( B\right) =\sum_{n}\varsigma\left( n,B\right)
|n\rangle\langle n|  \label{2.2}
\end{equation}
with respect to the standard transposition $\langle n|\widetilde{A}%
|m\rangle=\langle m|A|n\rangle$ in the eigenbasis of $\rho$.

Actually Ohya obtained the compound states $\omega_{d}$ as the result of the
composition 
\begin{equation*}
\omega_{d}\left( A\otimes B\right) =\omega_{o}\left( A\otimes\Lambda\left(
B\right) \right) ,
\end{equation*}
of quantum channels as normal unital CP maps $\Lambda:\mathcal{B}\rightarrow%
\mathcal{A}$ and the special, \emph{o-compound} states 
\begin{equation}
\omega_{o}\left( A\otimes B\right) =\sum_{n}\langle n|A|n\rangle p\left(
n\right) \langle n|B|n\rangle  \label{2.3}
\end{equation}
corresponding to the orthogonal decompositions 
\begin{equation}
\varpi_{o}\left( A\right) =\sum_{n}\langle n|A|n\rangle p\left( n\right)
|n\rangle\langle n|=\varpi_{o}^{\intercal}\left( A\right)  \label{2.4}
\end{equation}
such that $\varsigma_{n}\left( B\right) =\langle n|\Lambda\left( B\right)
|n\rangle$, $\sigma_{n}=\Lambda^{\intercal}\left( |n\rangle\langle n|\right) 
$, where $\left\langle B,\Lambda^{\intercal}\left( \rho\right) \right\rangle
_{\nu}=\mathrm{Tr}_{\mathcal{G}}\Lambda\left( B\right) \tilde{\rho}$.

Assuming that $\left\langle A,\rho\right\rangle =\mathrm{Tr}_{\mathcal{G}}A%
\tilde{\rho}$, we can extend this construction to any
discretely-decomposable algebra $\mathcal{A}=\widetilde{\mathcal{A}}$ on the
Hilbert sum $\mathcal{G}=\oplus\mathcal{G}_{i}$ with invariant components $%
\mathcal{G}_{i}$ under the standard complex conjugation $J$ in the
eigen-basis of the density operator $\tilde{\rho}=J\rho J=\rho$. In
particular, the von Neumann algebra $\mathcal{A}$ might be Abelian, as it is
in the case $\widetilde{A}=A$ for all $A\in\mathcal{A}$, e.g. when $\mathcal{%
A}=\widetilde{\mathcal{A}}$ is the diagonal algebra of pointwise
multiplications $Ag=ag=\widetilde{A}g$ by the bounded functions $n\mapsto
a\left( n\right) \in\mathbb{C}$ on the functional Hilbert space $\mathcal{G}%
=\ell^{2}\ni g$ with the standard complex conjugation $Jg=\bar{g}$. In this
case the densities $\rho\in\mathcal{A}_{\ast}$ are given by the summable
functions $p\in\ell^{1}$ with respect to the standard trace $\mu\left(
\rho\right) =\sum p\left( n\right) $, and any compound state has the
separable form with $\varrho_{n}\left( A\right) =a\left( n\right) $
corresponding to the Kronecker $\delta$-densities $\rho_{n}\simeq\delta_{n}$%
. The normal states on the $\mathcal{A}\simeq\ell^{\infty}$ are described by
the probability densities $p\left( n\right) \geq0$, $\sum p\left( n\right)
=1 $ with respect to the standard pairing 
\begin{equation*}
\left\langle A,\rho\right\rangle _{\mu}=\sum a\left( n\right) p\left(
n\right) ,\quad p\in\ell^{1},a\in\ell^{\infty}
\end{equation*}
of $\mathcal{A}_{\mu}=\mathcal{A}_{\ast}$ with the commutative algebra $%
\mathcal{A}$. Every normal compound state $\omega$ on $\mathcal{A}\otimes%
\mathcal{B}$ is defined by 
\begin{equation*}
\omega_{c}\left( A\otimes B\right) =\sum_{n}a\left( n\right) \left\langle
B,\sigma\left( n\right) \right\rangle _{\nu},
\end{equation*}
where $\sigma\left( n\right) =\sigma_{n}p\left( n\right) $ is the function
with positive values $\sigma\left( n\right) \in\mathcal{B}_{\intercal}$
normalized to the probability density $p\left( n\right) =\left\langle
I,\sigma\left( n\right) \right\rangle _{\nu}$. Thus all normal compound
states on $\ell^{\infty}\otimes\mathcal{B}$ are achieved by \emph{c-couplings%
} $\pi_{c}=\varpi_{c}^{\intercal}:\mathcal{B}\rightarrow\ell^{1}$ with $\pi
_{c}^{\intercal}=\varpi_{c}$ given by convex combinations of the primitive
CP (and TCP)\ maps $\varpi_{n}\left( a\right) =a\left( n\right) \sigma_{n}\in%
\mathcal{B}_{\ast}$, 
\begin{equation*}
\varpi_{c}\left( A\right) =\sum_{n}a\left( n\right) \sigma\left( n\right)
,\;\varpi_{c}^{\intercal}\left( B\right) =\sum_{n}\varsigma\left( n,B\right)
\delta_{n},
\end{equation*}
where $\varsigma\left( n,B\right) =\left\langle B,\sigma\left( n\right)
\right\rangle _{\nu}$. \ 

Note that any d-coupling can be regarded as quantum-classical c-coupling,
achieved by the identification $a\left( n\right) =\langle n|A|n\rangle$ of $%
\ell^{\infty}\ni a$ and the reduced diagonal algebra $\mathcal{A}%
^{0}=\left\{ \sum|n\rangle a\left( n\right) \langle n|:A\in\mathcal{A}%
\right\} $. This simply follows from the commutativity of the density
operators $\rho=\sum|n\rangle\langle n|p\left( n\right) $ for the induced
states $\varrho\left( A\right) =\omega_{d}\left( A\otimes I\right) $
identified with $p\in\ell^{1}$ .

In the case $\mathcal{A}=\mathcal{L}\left( \mathcal{G}\right) $\ and pure
elementary states $\omega_{n}$ described by probability amplitudes $%
\upsilon_{n}=\chi_{n}\otimes\psi_{n}$, where $\tilde{\chi}_{n}\equiv|\chi
_{n}\rangle\in\mathcal{G}$, $\tilde{\psi}_{n}\equiv|\psi_{n}\rangle \in%
\mathcal{H}$, we have density operators $\rho_{n}=$ $\chi_{n}^{\dagger}%
\chi_{n}$ and $\sigma_{n}=\psi_{n}^{\dagger}\psi_{n}$ of rank one. The total
compound amplitude is obviously $\upsilon=\sum|n\rangle\upsilon\left(
n\right) $, where $\upsilon\left( n\right) =\chi_{n}\otimes\psi_{n}p\left(
n\right) ^{1/2}$are the amplitude operators $\mathcal{G}\otimes \mathcal{H}%
\rightarrow\ell^{2}$ satisfying the orthogonality relations 
\begin{equation*}
\upsilon\left( n\right) ^{\dagger}\upsilon\left( m\right)
=\rho_{n}\otimes\sigma_{n}p\left( n\right) \delta_{n}^{m}
\end{equation*}
corresponding to the decomposition $\upsilon^{\dagger}\upsilon=\sum\rho
_{n}\otimes\sigma_{n}p\left( n\right) $. The ``entangling'' operator for the
separable state $\varkappa$ can be chosen as either as $\varkappa
=\sum|n\rangle\varkappa\left( n\right) $ or as $\varkappa=\sum
\varkappa\left( n\right) \langle n|$ or even as $\varkappa=\sum
|n\rangle\varkappa\left( n\right) \langle n|$ with $\varkappa\left( n\right)
=\chi_{n}\otimes\tilde{\psi}\left( n\right) $, where $\tilde{\psi }%
_{n}\left( n\right) =\tilde{\psi}_{n}p\left( n\right) ^{1/2}$. In particular
a d-entangling operator $\varkappa$ corresponding to d-encodings (\ref{2.2})
is diagonal$,$ $\varkappa=\sum|n\rangle\tilde{\psi}\left( n\right) \langle
n| $ on $\mathcal{G}=\ell^{2}$, corresponding to the orthogonal $\tilde{\chi}%
_{n}=|n\rangle$. Thus, we have proved the Theorem 2 below in the case of
pure states $\varsigma_{n}$ and $\varrho_{n}$. But, before formulating this
theorem in a natural generality let us introduce the following notations.

The general c-compound states on $\mathcal{A}\otimes \mathcal{B}$ are
defined as integral convex combinations 
\begin{equation*}
\omega \left( A\otimes B\right) =\int \varrho _{x}\left( A\right) \varsigma
_{x}\left( B\right) p\left( \mathrm{d}x\right)
\end{equation*}
given by a probability distribution $p$\textrm{\ }on the product-states $%
\varrho _{x}\otimes \varsigma _{x}$. Such compound states are achieved by
convex combinations of the primitive CP (and TCP)\ maps $\pi _{x}^{\symbol{%
126}}=\varpi _{x}^{\intercal }$ with $\varpi _{x}\left( A\right) =\varrho
_{x}\left( A\right) \sigma _{x}$: 
\begin{equation}
\varpi _{c}\left( A\right) =\int \varrho _{x}\left( A\right) \sigma
_{x}p\left( \mathrm{d}x\right) ,\;\varpi _{c}^{\intercal }\left( B\right)
=\int \varsigma _{x}\left( B\right) \rho _{x}p\left( \mathrm{d}x\right) .
\label{2.5}
\end{equation}
This is always the case when the von Neumann algebra $\mathcal{A}$ is
Abelian, and thus can be identified with the diagonal algebra of
multiplications $\left( Ag\right) \left( x\right) =a\left( x\right) g\left(
x\right) $ by the functions $a\in L_{\mu }^{\infty }$ on the functional
Hilbert space $\mathcal{G}=L_{\mu }^{2}$ with respect to a (not necessarily
finite) measure $\mu $ on $X$. It defines trace $\mu $ on $\mathcal{A}_{\mu
}\simeq L_{\mu }^{1}\cap L_{\mu }^{\infty }$ as the integral $\mu \left(
\rho \right) =\int p\left( x\right) \mu \left( \mathrm{d}x\right) $ for the
bounded multiplication densities $\left( \rho g\right) \left( x\right)
=p\left( x\right) g\left( x\right) $. The normal states on $\mathcal{A}$ are
given by the probability densities $p\in L_{\mu }^{1}$ with respect to the
standard pairing 
\begin{equation*}
\left\langle A,\rho \right\rangle _{\mu }=\int a\left( x\right) p\left(
x\right) \mu \left( \mathrm{d}x\right) ,\quad p\in L_{\mu }^{1},a\in L_{\mu
}^{\infty }
\end{equation*}
of $\mathcal{A}_{\ast }=\mathcal{A}_{\intercal }\simeq L_{\mu }^{1}$ and $%
\mathcal{A}=\widetilde{\mathcal{A}}\simeq L_{\mu }^{\infty }$ corresponding
to the trivial transposition $\tilde{a}=a$. Any normal compound state $%
\omega $ on $\mathcal{A}\otimes \mathcal{B}\simeq L_{\mu }^{\infty }\left(
X\rightarrow \mathcal{B}\right) $ is the c-compound state, defined on the
diagonal algebra $\mathcal{A}$ by 
\begin{equation}
\omega _{d}\left( A\otimes B\right) =\int a\left( x\right) \varsigma \left(
x,B\right) \mu \left( \mathrm{d}x\right) ,  \label{2.6}
\end{equation}
where $\varsigma \left( x,B\right) =\left\langle B,\sigma \left( x\right)
\right\rangle _{\nu }$ is an absolutely integrable function with density
operator values $\sigma \left( x\right) =\sigma _{x}p\left( x\right) $
normalized to the probability density $p\left( x\right) =\left\langle
I,\sigma \left( x\right) \right\rangle _{\nu }=\varsigma \left( x,I\right) $%
. It corresponds to d-couplings $\pi _{d}=\varpi _{d}^{\intercal }=\pi _{d}^{%
\symbol{126}}$ with $\pi _{d}^{\intercal }=\varpi _{d}$ decomposing into $%
\varpi \left( x,A\right) =a\left( x\right) \sigma \left( x\right) $: 
\begin{equation}
\varpi _{d}\left( A\right) =\int a\left( x\right) \sigma \left( x\right) \mu
\left( \mathrm{d}x\right) ,\;\varpi _{d}^{\intercal }\left( B\right) =\int
\varsigma \left( x,B\right) \delta _{x}\mu \left( \mathrm{d}x\right) ,
\label{2.8}
\end{equation}
where $\delta _{x}$ is the (generalized) density operator of the Dirac state 
$\varrho _{x}\left( A\right) =\left\langle A,\delta _{x}\right\rangle _{\mu
}=a\left( x\right) $ on the diagonal algebra $\mathcal{A}$.

\begin{theorem}
Let $\omega _{c}:\mathcal{A}\otimes \mathcal{B}\rightarrow \mathbb{C}$ be a
normal c-compound state given as 
\begin{equation}
\omega _{c}\left( A\otimes B\right) =\int \mu _{x}\left( \tilde{\chi}_{\cdot
}^{\dagger }A\tilde{\chi}_{\cdot }\right) \nu _{x}\left( \tilde{\psi}_{\cdot
}^{\dagger }B\tilde{\psi}_{\cdot }\right) p\left( \mathrm{d}x\right) ,
\label{2.7}
\end{equation}
where $\chi _{x}:\mathcal{G}\rightarrow \mathcal{E}_{x}$, $\psi _{x}:%
\mathcal{H}\rightarrow \mathcal{F}_{x}$ are linear operators having bounded
transpose $\tilde{\chi}_{\cdot }=J\chi _{\cdot }^{\dagger }J_{\cdot }$ $%
\tilde{\psi}_{\cdot }=J\psi _{\cdot }^{\dagger }J_{\cdot }$ on Hilbert
spaces $\mathcal{E}_{\cdot }=\int^{\oplus }\mathcal{E}_{x}p\left( \mathrm{d}%
x\right) $, $\mathcal{F}_{\cdot }=\int^{\oplus }\mathcal{F}_{x}p\left( 
\mathrm{d}x\right) $ with respect to pointwise involution $J_{\cdot
}=J_{\cdot }^{\dagger }$. We also assume\ that 
\begin{equation*}
\chi _{x}^{\dagger }\chi _{x}\in \widetilde{\mathcal{A}},\psi _{x}^{\dagger
}\psi _{x}\in \widetilde{\mathcal{B}},\quad \mu _{x}\left( \tilde{\chi}%
_{\cdot }^{\dagger }\tilde{\chi}_{\cdot }\right) =1=\nu _{x}\left( \tilde{%
\psi}_{\cdot }^{\dagger }\tilde{\psi}_{\cdot }\right)
\end{equation*}
with respect to the traces 
\begin{equation}
\mu _{x}\left( \tilde{\chi}_{\cdot }^{\dagger }\tilde{\chi}_{\cdot }\right)
=\left\langle I,\chi _{x}^{\dagger }\chi _{x}\right\rangle _{\mu },\quad \nu
_{x}\left( \tilde{\psi}_{\cdot }^{\dagger }\tilde{\psi}_{\cdot }\right)
=\left\langle I,\psi _{x}^{\dagger }\psi _{x}\right\rangle _{\nu }.
\label{2.9}
\end{equation}
Then this state is achieved by a decomposable entangling operator $\varkappa
=\int^{\oplus }\chi _{x}\otimes \tilde{\psi}_{x}p\left( \mathrm{d}x\right) $
defining c-entanglement (\ref{2.5}) with 
\begin{equation}
\varrho _{x}\left( A\right) =\mu _{x}\left( \tilde{\chi}_{\cdot }^{\dagger }A%
\tilde{\chi}_{\cdot }\right) ,\;\varsigma _{x}\left( B\right) =\nu
_{x}\left( \tilde{\psi}_{\cdot }^{\dagger }A\tilde{\psi}_{\cdot }\right) ,
\label{2.10}
\end{equation}
corresponding to the probability densities $\rho _{x}=\chi _{x}^{\dagger
}\chi _{x}$, $\sigma _{x}=\psi _{x}^{\dagger }\psi _{x}$. In particular,
every d-compound state (\ref{2.6}) corresponding to $p\left( \mathrm{d}%
x\right) =p\left( x\right) \mu \left( \mathrm{d}x\right) $ with the Abelian
algebra $\mathcal{A}$ can be achieved by the orthogonal sum of entangling
operators $\varkappa _{x}=\delta _{x}\otimes \tilde{\psi}_{x}$ defining
d-entanglement (\ref{2.8}) with 
\begin{equation*}
\sigma \left( x\right) =\psi _{x}^{\dagger }\psi _{x}p\left( x\right) ,\quad
\varsigma \left( x,B\right) =\nu _{x}\left( \tilde{\psi}_{\cdot }^{\dagger }A%
\tilde{\psi}_{\cdot }\right) p\left( x\right) .
\end{equation*}
\end{theorem}

\proof%
%
The amplitude operator $\upsilon=\int^{\otimes}\upsilon_{x}p\left( \mathrm{d}%
x\right) $ corresponding to c-compound state (\ref{2.7}) is defined as the
orthogonal sum of $\upsilon_{x}=\chi_{x}\otimes\psi_{x}$ on $\mathcal{G}%
\otimes\mathcal{H}$ into $\int^{\oplus}\mathcal{E}_{x}\otimes\mathcal{F}%
_{x}p\left( \mathrm{d}x\right) $. Without loss of generality we can assume
that $\mathcal{E}_{x}=\mathcal{G}_{\rho}$, $\mathcal{F}_{x}=\mathcal{H}%
_{\sigma}$ and $\upsilon_{x}^{\dagger}=\upsilon_{x}\left( E_{\rho}\otimes
E_{\sigma}\right) $ because the support $\left( \mathcal{G}\otimes\mathcal{H}%
\right) _{\upsilon_{x}^{\dagger }\upsilon_{x}}=\mathrm{ran}%
\upsilon_{x}^{\dagger}$ for 
\begin{equation*}
\upsilon_{x}^{\dagger}\upsilon_{x}=\chi_{x}^{\dagger}\chi_{x}\otimes\psi
_{x}^{\dagger}\psi_{x}=\rho_{x}\otimes\sigma_{x}
\end{equation*}
is in $\mathcal{G}_{\rho}\otimes\mathcal{H}_{\sigma}$. Due to $\chi
_{x}^{\dagger}\chi_{x}\in\widetilde{\mathcal{A}^{\prime}}^{\prime}$, $\psi
_{x}^{\dagger}\psi_{x}\in\widetilde{\mathcal{B}^{\prime}}^{\prime}$ for
almost all $x$, the operators $\chi_{x}$ and $\psi_{x}$ commute with $A\in 
\widetilde{\mathcal{A}^{\prime}}$ and $B\in\widetilde{\mathcal{B}^{\prime}}$
respectively, and $\tilde{\psi}_{x}$ commutes with $B\in\mathcal{B}^{\prime}$
for almost all $x$. Thus, 
\begin{equation*}
\tilde{\chi}_{x}^{\dagger}A\tilde{\chi}_{x}\in\mathcal{A},\quad\tilde{\psi }%
_{x}B\tilde{\psi}_{x}\in\mathcal{B}
\end{equation*}
which defines the traces (\ref{2.9}) on $L_{p}^{\infty}\otimes\mathcal{A}$
and $L_{p}^{\infty}\otimes\mathcal{B}$ for almost all $x$. The rest of the
proof is a repetition of the proof of Theorem 1 for each $x$, with the
addition that $\varkappa_{x}$ is the product $\upsilon_{x}^{\prime}=\chi_{x}%
\otimes \tilde{\psi}_{x}$ for each $x$. The total entangling operator $%
\varkappa :\mathcal{G}\otimes\mathcal{F}_{\cdot}\rightarrow\mathcal{E}%
_{\cdot}\otimes\mathcal{H}$ acts componentwise as $\varkappa_{x}\left( \zeta
\otimes\eta_{\cdot}\right) =\chi_{x}\zeta\otimes\tilde{\psi}_{x}\eta_{x}$.

In the case of d-compound state (\ref{2.6}) one should take $\mathcal{G}%
=L_{\mu}^{2}$, $\mathcal{E}_{x}=\mathbb{C}$, and $\chi_{x}g=g\left( x\right) 
$. Thus the entangling operator in this case is given as 
\begin{equation*}
\varkappa\left( g\otimes\eta_{\cdot}\right) =\int^{\otimes}g\left( x\right) 
\tilde{\psi}_{x}\eta_{x}\mu\left( \mathrm{d}x\right) ,\quad\forall g\in
L_{\mu}^{2},\eta_{\cdot}=\int^{\oplus}\eta_{x}\mu\left( \mathrm{d}x\right)
\in\mathcal{F}_{\cdot}.
\end{equation*}
\endproof%
%

Note that c-entanglements $\varpi _{c}$ in (\ref{2.5}) are both CP and TCP
and thus are not true quantum entanglements. The map $\varpi _{c}:\mathcal{A}%
\rightarrow \mathcal{B}_{\intercal }$ with Abelian algebra $\mathcal{A}$ in (%
\ref{2.8}) is described by a $\mathcal{B}_{\intercal }$-valued measure $%
\sigma \left( \mathrm{d}x\right) =\sigma \left( x\right) \mu \left( \mathrm{d%
}x\right) $ normalized to the input probability measure as $p\left( \mathrm{d%
}x\right) =\left\langle I,\sigma \left( \mathrm{d}x\right) \right\rangle
_{\nu }$. This gives the concise form for the description of random
classical-quantum state correspondences $x\mapsto \sigma _{x}$ with the
given probability measure $p$, called \emph{encodings} of $\sigma =\int
\sigma \left( \mathrm{d}x\right) $.

\begin{definition}
Let both algebras $\mathcal{A}$ and $\mathcal{B}$ be non-Abelian. The map $%
\varpi :\mathcal{A}\rightarrow \mathcal{B}_{\intercal }$ is called a \emph{%
c-encoding} of $\left( \mathcal{B},\varsigma \right) $ if it is a convex
combination of the primitive maps $\sigma _{n}\varrho _{n}$ given by the
probability densities $\sigma _{n}\in \mathcal{B}_{\intercal }$ and normal
states $\varrho _{n}:\mathcal{A}\rightarrow \mathbb{C}$. It is called \emph{%
d-encoding} if it has the diagonalizing form (\ref{2.2}) on $\mathcal{A}$,
and it is called \emph{o-encoding} if all density operators $\sigma _{n}$
are mutually orthogonal: $\sigma _{m}\sigma _{n}=0$ for all $m\neq n$ as in (%
\ref{2.4}). The entanglement which is described by non-separable CP map $%
\varpi :\mathcal{A}\rightarrow \mathcal{B}_{\intercal }$ will be called 
\emph{q-encoding}.
\end{definition}

Note that due to the commutativity of the operators $A\otimes I$ with $%
I\otimes B$ on $\mathcal{G}\otimes\mathcal{H}$, one can treat the encodings
as nondemolition measurements \cite{Bel94} in $\mathcal{A}$ with respect to $%
\mathcal{B}$. The corresponding compound state is the state prepared for
such measurements on the input $\mathcal{G}$. It coincides with the mixture
of the states corresponding to those after the measurement, without reading
the message sent. The set of all d-encodings for a Schatten decomposition of
the input state $\rho$ on $\mathcal{A}$ is obviously convex with the extreme
points given by the pure output states $\varsigma_{n}$ on $\mathcal{B}$,
corresponding to the not necessarily orthogonal (not Schatten)
decompositions $\sigma=\sum\sigma\left( n\right) $ into the one-dimensional
density operators $\sigma\left( n\right) =p\left( n\right) \sigma_{n}.$

The Schatten decompositions $\sigma=\sum_{n}q\left( n\right) \sigma_{n}$
correspond to o-encodings, the extreme d-encodings $\sigma_{n}=\eta_{n}%
\eta_{n}^{\dagger}$, $p\left( n\right) =q\left( n\right) $ characterized by
the orthogonality $\sigma_{m}\sigma_{n}=0$, $m\neq n$ . For each Schatten
decomposition of $\sigma$ they form a convex subset of d-encodings with
mixed commuting $\sigma_{n}$ .

\section{Quantum versus von Neumann Entropy.}

As we have seen in the previous section, the encodings $\varpi:\mathcal{A}%
\rightarrow\mathcal{B}_{\intercal}$, which are described in (\ref{2.8})
usually with a discrete Abelian $\mathcal{A}$, correspond to the case (\ref
{2.2}) when the general entanglement (\ref{1.6}) is \emph{d-encoding}, with
the diagonal coupling $\pi=\varpi^{\intercal}$ in the eigen-representation
of a discrete probability density $\rho$ on non-Abelian $\mathcal{A}$. The
true quantum entanglements with non-Abelian $\mathcal{A}$ cannot be achieved
by d-, or more generally, c-encodings even in the case of discrete $\mathcal{%
A}$. The nonseparable, true entangled states $\omega$ called in \cite{Ohy89} 
\emph{q-compound} states, can be achieved by \emph{q-encodings}, the
quantum-quantum nonseparable correspondences (\ref{1.5}) which are not
diagonal in the eigen-representation of $\rho$.

As we shall prove in this section, the self-dual standard true entanglement $%
\varpi_{q}=\varpi_{q}^{\intercal}$ to the probe system $\left( \mathcal{A}%
^{0},\varrho_{0}\right) =\left( \widetilde{\mathcal{B}},\tilde{\varsigma}%
\right) $, which is defined in (\ref{1.9}), is the most informative for a
quantum system $\left( \mathcal{B},\varsigma\right) $, in the sense that it
achieves the maximal mutual information in the coupled system $\left( 
\mathcal{A}\otimes\mathcal{B},\omega\right) $ when $\omega=\omega_{q}$ is
given in (\ref{1.10}).

Let us consider entangled mutual information and quantum entropies of states
by means of the above three types of compound states. To define the quantum
mutual entropy we need to apply a quantum version of the relative entropy to
compound states on the algebra $\mathcal{M}=\mathcal{A}\otimes\mathcal{B}$,
called also the information divergency of the state $\omega$ with respect to
a reference state $\varphi$ on $\mathcal{M}$. The relative entropy was
defined in \cite{Lin73, Ara76, Ume}, even for the most general von Neumann
algebra $\mathcal{M}$, but for our purposes we need the following explicit
formulation.

Let $\mathcal{M}$ be a semi-finite algebra with normal states $\omega$ and $%
\varphi$ having the density operator $\upsilon^{\dagger}\upsilon$ and $%
\phi\in\widetilde{\mathcal{M}}$ with respect to the pairing 
\begin{equation*}
\left\langle M,\upsilon^{\dagger}\upsilon\right\rangle =\tau\left( \upsilon%
\widetilde{M}\upsilon^{\dagger}\right) ,\quad M\in\mathcal{M}%
,\upsilon\upsilon^{\dagger}\in\widetilde{\mathcal{M}}
\end{equation*}
given by a normal faithful trace $\tau$ on the transposed algebra $%
\widetilde{\mathcal{M}}=J\mathcal{M}J$ (not necessary decomposable as $\tau=%
\tilde{\mu}\otimes\tilde{\nu}$ in (\ref{1.3}) in the case of $\mathcal{M}=%
\mathcal{A}\otimes\mathcal{B}$). Then the \emph{relative entropy} $\mathsf{R}%
\left( \omega;\varphi\right) $ of the state $\omega$ with respect to $%
\varphi $ is given by the formula 
\begin{equation}
\mathsf{R}\left( \omega:\varphi\right) =\tau\left( \upsilon\left(
\ln\upsilon^{\dagger}\upsilon-\ln\phi\right) \upsilon^{\dagger}\right)
=\tau\left( \omega\left( \ln\omega-\ln\phi\right) \right) .  \label{4.1}
\end{equation}
(For notational simplicity here and below we identify the state $\omega$
with its density operator $\upsilon^{\dagger}\upsilon$). It has a positive
value $\mathsf{R}\left( \omega:\varphi\right) \in\lbrack0,\infty]$ if the
states are equally normalized, say (as usual) $\tau\left( \mathrm{\,}%
\omega\right) =1=\tau\left( \phi\right) $, and it can be finite only if the
state $\omega$ is absolutely continuous with respect to the reference state $%
\varphi$, i.e. iff $\omega\left( E\right) =0$ for the maximal
null-orthoprojector $E\in\mathcal{M}$, $E\phi=0$. This definition does not
depend on the choice of the semi-finite trace $\tau$, and it can be extended
also to the arbitrary normal $\omega$ and $\varphi$ with unbounded
self-adjoint density operators $\upsilon^{\dagger}\upsilon$ and $\phi$.

The most important property of the information divergence $\mathsf{R}$ is
its monotonicity property \cite{Lin73, Uhl}, i.e. nonincrease in the
divergency $\mathsf{R}\left( \omega_{0}:\varphi_{0}\right) $ after the
application of the pre-dual of a normal completely positive unital map $%
\mathrm{K}:\mathcal{M}\rightarrow\mathcal{M}^{0}$ to the states $\omega_{0} $
and $\varphi_{0}$ on a von Neumann algebra $\mathcal{M}^{0}$: 
\begin{equation}
\left( \omega=\omega_{0}\mathrm{K},\varphi=\varphi_{0}\mathrm{K}\right)
\quad\Rightarrow\quad\mathsf{R}\left( \omega:\varphi\right) \leq \mathsf{R}%
\left( \omega_{0}:\varphi_{0}\right) .  \label{4.2}
\end{equation}

The \emph{mutual information} $\mathsf{I}\left( \pi \right) =\mathsf{I}%
\left( \pi ^{\ast }\right) $ in a compound state $\omega $, achieved by a
coupling $\pi :\mathcal{B}\rightarrow \mathcal{A}_{\ast }$, or by $\pi
^{\ast }:\mathcal{A}\rightarrow \mathcal{B}_{\ast }$ with the marginals 
\begin{equation*}
\varrho \left( A\right) =\omega \left( A\otimes I\right) =\left\langle
A,\rho \right\rangle _{\mu },\;\varsigma \left( B\right) =\omega \left(
I\otimes B\right) =\left\langle B,\sigma \right\rangle _{\nu },
\end{equation*}
is defined by the relative entropy 
\begin{equation}
\mathsf{I}\left( \pi \right) =\tau \left( \mathrm{\,}\omega \left( \ln
\omega -\ln \left( \rho \otimes I\right) -\ln \left( I\otimes \sigma \right)
\right) \right) =\mathsf{R}\left( \omega :\varrho \otimes \varsigma \right)
\label{4.3}
\end{equation}
of the state $\omega $ on $\mathcal{M}=\mathcal{A}\otimes \mathcal{B}$ with
respect to the product state $\varphi =\varrho \otimes \varsigma $ for $\tau
=\tilde{\mu}\otimes \tilde{\nu}$. This quantity, generalizing the classical
mutual information corresponding to the case of Abelian $\mathcal{A}$, $%
\mathcal{B}$, describes an information gain in a quantum system $\left( 
\mathcal{A},\varrho \right) $ via the entanglement $\varpi ^{\intercal }=\pi
^{\symbol{126}}$, or in $\left( \mathcal{B},\varsigma \right) $ via an
entanglement $\varpi :\mathcal{A}\rightarrow \mathcal{B}_{\intercal }$. It
is naturally treated as a measure of the strength of the generalized
entanglement having zero value only for completely disentangled states $%
\omega =\varrho \otimes \varsigma $.

\begin{proposition}
Let $\left( \mathcal{A}^{0},\mu _{0}\right) $ be a quantum system with a
normal faithful semifinite trace, and $\pi _{0}:\mathcal{A}^{0}\rightarrow 
\mathcal{B}_{\ast }$ be a normal coupling of the state $\varrho _{0}=\nu
\circ \pi _{0}$ on $\mathcal{A}^{0}$ to $\varsigma =\mu \circ \pi $,
defining an entanglement $\varpi =\pi ^{\ast \symbol{126}}$ of $\left( 
\mathcal{A},\varrho \right) $ to $\left( \mathcal{B},\varsigma \right) $ by
the composition $\pi ^{\ast }=\pi _{0}\mathrm{K}$ with a normal completely
positive unital map $\mathrm{K}:\mathcal{A}\rightarrow \mathcal{A}^{0}$.
Then $\mathsf{I}\left( \pi \right) \leq \mathsf{I}\left( \pi ^{0}\right) $,
where $\pi ^{0}=\pi _{0}^{\ast }$. In particular, for each normal c-coupling
given by (\ref{2.5}) such as $\pi ^{\symbol{126}}=\varpi _{c}^{\intercal }$
there exists a not less informative d-coupling $\pi ^{0}=\varpi
_{0}^{\intercal }$ with Abelian $\mathcal{A}^{0}$ corresponding to the
encoding $\varpi _{0}=\pi _{0}^{\symbol{126}}\ $of $\left( \mathcal{B}%
,\varsigma \right) $, and the standard q-coupling $\pi ^{0}=\pi _{q}$, $\pi
_{q}\left( B\right) =\sigma ^{1/2}\tilde{B}\sigma ^{1/2}$ to $\varrho _{0}=%
\tilde{\varsigma}$ on $\mathcal{A}^{0}=\widetilde{\mathcal{B}}$ is the
maximal coupling in this sense.
\end{proposition}

\proof%
%
The first follows from the monotonicity property (\ref{4.2}) applied to the
extension $\mathrm{K}\left( A\otimes B\right) =\mathrm{K}\left( A\right)
\otimes B$ of the CP map $\mathrm{K}$ from $\mathcal{A}\rightarrow \mathcal{A%
}^{0}$ to $\mathcal{A}\otimes\mathcal{B}\rightarrow\mathcal{A}^{0}\otimes%
\mathcal{B}$. The compound state $\omega_{0}\left( \mathrm{K}\otimes\mathrm{I%
}\right) $ ($\mathrm{I}$ denotes the identity map $\mathcal{B}\rightarrow%
\mathcal{B}$) is achieved by the entanglement $\varpi=\varpi_{0}\mathrm{K}$
and $\varphi_{0}\left( \mathrm{K}\otimes\mathrm{I}\right)
=\varrho\otimes\varsigma$ , $\varrho=\varrho _{0}\mathrm{K}$ corresponding
to $\varphi_{0}=\varrho_{0}\otimes\varsigma$. It corresponds to the coupling 
$\pi=\mathrm{K}^{\ast}\pi_{0}$ which is defined by $\mathrm{K}^{\ast}:%
\mathcal{A}_{\ast}^{0}\rightarrow\mathcal{A}_{\ast}$ as $\mathrm{K}^{\ast}%
\tilde{\rho}_{0}=J\left( \mathrm{K}^{\intercal}\rho _{0}\right) ^{\dagger}J$%
, where 
\begin{equation*}
\left\langle A,\mathrm{K}^{\intercal}\rho_{0}\right\rangle
_{\mu}=\left\langle \mathrm{K}A,\rho_{0}\right\rangle
_{\mu_{0}},\quad\forall A\in\mathcal{A},\rho_{0}\in\mathcal{A}%
_{\intercal}^{0}.
\end{equation*}

This monotonicity property proves, in particular, that for any separable
compound state (\ref{2.7}) on $\mathcal{A}\otimes\mathcal{B}$, which is
prepared by the c-entanglement $\pi_{c}^{\symbol{126}}=\varpi_{c}^{%
\intercal} $, there exists a d-entanglement $\varpi_{0}^{\intercal}=\pi_{0}$
with $\left( \mathcal{A}^{0},\varrho_{0}\right) $ having the same, or even
larger information gain (\ref{4.3}). One can even take a classical system $%
\left( \mathcal{A}^{0},\varrho_{0}\right) $, say the diagonal sublalgebra $%
\mathcal{A}^{0}\simeq L_{p}^{\infty}$ on $\mathcal{G}_{0}=$ $L_{p}^{2}$ with
the state $\varrho_{0}$, induced by the measure $\mu=p$, and consider the
classical-quantum correspondence (encoding) 
\begin{equation*}
\varpi_{0}\left( A^{0}\right) =\int a\left( x\right) \sigma_{x}p\left( 
\mathrm{d}x\right) ,\quad A^{0}=\int^{\oplus}a\left( x\right) p\left( 
\mathrm{d}x\right) ,a\in L_{p}^{\infty}
\end{equation*}
assigning the states $\varsigma_{x}\left( B\right) =\left\langle
B,\sigma_{x}\right\rangle _{\nu}$ to the letters $x$ with the probabilities $%
p\left( \mathrm{d}x\right) $. In this case the state $\varrho$ is described
by the density $\rho=I$, the multiplication by identity function in $%
L_{p}^{2}$, $\omega$ is multiplication by $\sigma_{\cdot}=\left( \sigma
_{x}\right) $ in $L_{p}^{2}\otimes\mathcal{H}$ and the mutual information (%
\ref{4.3}) is given as 
\begin{equation}
\mathsf{I}\left( \pi^{0}\right) =\int\mathrm{\,}\tilde{\nu}_{x}\left(
\sigma_{x}\left( \ln\sigma_{x}-\ln\sigma\right) \right) p\left( \mathrm{d}%
x\right) =\mathsf{S}\left( \sigma\right) -\int\mathsf{S}\left(
\sigma_{x}\right) p\left( \mathrm{d}x\right) ,  \label{4.4}
\end{equation}
where $\mathsf{S}\left( \sigma\right) =-\tilde{\nu}\left( \sigma\ln
\sigma\right) $. The achieved information gain $\mathsf{I}\left( \pi
^{0}\right) $ is larger than $\mathsf{I}\left( \pi\right) $ corresponding to 
$\omega=\int\rho_{x}\otimes\sigma_{x}p\left( \mathrm{d}x\right) $ because
the c-entanglement $\varpi_{c}$ in (\ref{2.5}) is represented as the
composition $\varpi_{0}\mathrm{K}$ of the encoding $\varpi_{0}:\mathcal{A}%
^{0}\rightarrow\mathcal{B}_{\intercal}$ with the CP map 
\begin{equation*}
\mathrm{K}\left( A\right) =\int^{\oplus}\varrho_{x}\left( A\right) p\left( 
\mathrm{d}x\right) ,\quad A\in\mathcal{A}
\end{equation*}
given by $a\left( x\right) =\varrho_{x}\left( A\right) $ for each $A\in%
\mathcal{A}$. Hence 
\begin{equation*}
\pi^{\ast}\left( A\right) =\widetilde{\varpi\left( A\right) }=\widetilde{%
\varpi_{0}\mathrm{K}A}=\pi_{0}\left( \mathrm{K}A\right) ,\quad\forall A\in%
\mathcal{A}
\end{equation*}
where $\pi_{0}=\varpi_{0}^{\symbol{126}}$, and thus $\mathsf{I}\left( \pi
^{0}\right) \geq\mathsf{I}\left( \mathrm{K}^{\ast}\pi^{0}\right) =\mathsf{I}%
\left( \pi\right) $, where $\pi^{0}=\pi_{0}^{\ast}=\varpi _{0}^{\intercal}$.

The inequality (\ref{4.2}) can also be applied to the standard entanglement
corresponding to the compound state (\ref{1.10}) on $\widetilde{\mathcal{B}}%
\otimes\mathcal{B}$. Indeed, any normal entanglement $\varpi\left( A\right)
=\mu\left( \tilde{\varkappa}^{\dagger}\left( A\otimes I\right) \tilde{%
\varkappa}\right) $, on $\mathcal{A}$ into $\mathcal{B}_{\intercal}$
described by a CP map $\mathcal{A}\rightarrow\widetilde{\mathcal{B}_{\nu}}$,
can be decomposed as 
\begin{equation*}
\mu\left( \tilde{\varkappa}^{\dagger}\left( A\otimes I\right) \tilde{%
\varkappa}\right) =\sigma^{1/2}\mu\left( X^{\dagger}\left( A\otimes I\right)
X\right) \sigma^{1/2}=\varpi_{0}\left( \mathrm{K}A\right) ,
\end{equation*}
where $\mathrm{K}A=\mu\left( X^{\dagger}\left( A\otimes I\right) X\right) $
is a normal unital CP map $\mathcal{A}\rightarrow\widetilde{\mathcal{B}}$.
It is uniquely given by an operator $X:\mathcal{E}\otimes\mathcal{H}%
\rightarrow\mathcal{G}\otimes\mathcal{F}$ \ with $\mathcal{E}=\mathcal{G}%
_{\rho}$, $\mathcal{H}=\mathcal{F}_{\sigma}$ satisfying the condition $%
X\left( I\otimes\sigma\right) ^{1/2}=\tilde{\varkappa}$, and thus $X\in%
\mathcal{A}\otimes\mathcal{B}^{\prime}$ due to the commutativity of $\tilde{%
\varkappa}$ with $\mathcal{A}^{\prime}\otimes\mathcal{B}$ and $\sigma$ with $%
\mathcal{B}$. Moreover, the partial trace $\mu$ of $X^{\dagger}X$ is
well-defined\ by $\mu\left( \tilde{\varkappa}^{\dagger}\tilde{\varkappa }%
\right) =\sigma$ as $\mu\left( X^{\dagger}X\right) =I$. Thus $%
\varpi=\varpi_{q}\mathrm{K}$ and $\pi=\mathrm{K}^{\ast}\pi_{q}$, where $%
\mathrm{K}$ is a normal unital CP map $\mathcal{A}\rightarrow\widetilde {%
\mathcal{B}}$, and $\mathrm{K}^{\ast}:\mathcal{B}_{\intercal}=\widetilde {%
\mathcal{B}}_{\ast}\rightarrow\mathcal{A}_{\ast}$. Hence the standard
entanglement (coupling) (\ref{1.9}) corresponds to the maximal mutual
information, $\mathsf{I}\left( \pi_{q}\right) \geq\mathsf{I}\left( \mathrm{K}%
^{\ast}\pi_{q}\right) =\mathsf{I}\left( \pi\right) $.%
\endproof%
%

Note that the mutual information (\ref{4.3}) is written as 
\begin{equation*}
\mathsf{I}\left( \pi\right) =\mathsf{S}\left( \rho\right) +\mathsf{S}\left(
\sigma\right) -\mathsf{S}\left( \omega/\varphi\right) ,
\end{equation*}
where $\varphi=\mu\otimes\nu$, $\mathsf{S}\left( \rho\right) =\mathsf{S}%
\left( \varrho/\mu\right) $, $\mathsf{S}\left( \sigma\right) =\mathsf{S}%
\left( \varsigma/\nu\right) $ and 
\begin{equation}
\mathsf{S}\left( \omega/\varphi\right) =-\tilde{\varphi}\left(
\upsilon\left( \ln\upsilon^{\dagger}\upsilon\right) \upsilon^{\dagger
}\right) \equiv-\tilde{\varphi}\left( \upsilon^{\dagger}\upsilon\ln
\upsilon^{\dagger}\upsilon\right)  \label{4.5}
\end{equation}
denotes the \emph{entropy of the density operator} $\upsilon^{\dagger}%
\upsilon\in\widetilde{\mathcal{M}}$\ of the state $\omega$ with respect to
the trace $\varphi$ on $\mathcal{M}$. Note that the entropy $\mathsf{S}%
\left( \omega/\varphi\right) $, coinciding with $-\mathsf{R}\left( \omega
:\varphi\right) $ (cf. with (\ref{4.1}) in the case $\tau=\tilde{\varphi}$),
is not in general positive, and may not even be bounded from below as a
function of $\omega$. However, in the case of irreducible $\mathcal{M}$ it
can always be made positive by the choice of the standard trace $\tau=%
\mathrm{Tr}$ on $\mathcal{M}$, in which case it is called the von Neumann
entropy of the state $\omega$ ($=\upsilon^{\dagger}\upsilon$), denoted
simply as $\mathsf{S}\left( \omega\right) $: 
\begin{equation}
\mathsf{S}\left( \omega/\tau\right) =-\mathrm{Tr}\omega\ln\omega \equiv%
\mathsf{S}\left( \omega\right) .  \label{4.6}
\end{equation}

In the following we shall assume that $\mathcal{B}$ is a discrete
decomposition of the irreducible $\mathcal{B}_{i}=\mathcal{L}\left( \mathcal{%
H}_{i}\right) =\widetilde{\mathcal{B}_{i}}$ with the trace $\nu=\mathrm{Tr}_{%
\mathcal{H}}=\tilde{\nu}$ induced on $\mathcal{B}_{\ast }=\mathcal{B}_{\tau}$%
. The entropy $\mathsf{S}\left( \sigma\right) =\mathsf{S}\left(
\varsigma/\nu\right) $ of the density operator $\sigma$ for the\ normal
state $\varsigma$ on $\mathcal{B}$ can be found in this case as the maximal
information $\mathsf{S}\left( \varsigma\right) =\sup \mathsf{I}\left(
\pi_{c}\right) $ achieved via all c-encodings $\varpi:\mathcal{A}\mapsto%
\mathcal{B}_{\tau}$ of the system $\left( \mathcal{B},\mathcal{\varsigma}%
\right) $ such that, $\varpi\left( I\right) =\sigma$, $\varpi^{\intercal}=%
\pi_{c}^{\symbol{126}}$. Indeed, as follows from the proposition above, it
is sufficient to find the maximum of $\mathsf{I}\left( \pi\right) $ over all
d-couplings $\pi=\varpi^{\intercal}$ mapping $\mathcal{B}$ into Abelian $%
\mathcal{A}$ with fixed $\varpi\left( I\right) =\sigma$, i.e. to find
maximum of (\ref{4.4}) under the condition $\int \sigma_{x}p\left( \mathrm{d}%
x\right) =\sigma$. Due to positivity of the \emph{d-conditional entropy} 
\begin{equation}
\mathsf{S}\left( \pi_{d}\right) =-\int\mathrm{Tr}\left( \sigma_{x}\ln
\sigma_{x}\right) p\left( \mathrm{d}x\right) =\int\mathsf{S}\left(
\sigma_{x}\right) p\left( \mathrm{d}x\right)  \label{4.7}
\end{equation}
the information $\mathsf{I}\left( \pi^{0}\right) =\mathsf{I}\left( \pi
_{d}\right) $ has the maximum $\mathsf{S}\left( \sigma\right) $ which is
achieved on an extreme d-coupling $\pi_{d}^{0}$ when almost all $\mathsf{S}%
\left( \sigma_{x}\right) $ are zero, i.e. when almost all $\sigma_{x}$ are
one-dimensional projectors $\sigma_{x}^{0}=P_{x}$ corresponding to pure
states $\varsigma_{x}$. One can take for example, the maximal Abelian
subalgebra $\mathcal{A}^{0}\subseteq\mathcal{B}$ generated by $%
P_{n}=|n\rangle\langle n|\in\mathcal{B}$ for a Schatten decomposition $%
\sigma=\sum_{n}|n\rangle \langle n|p\left( n\right) $ of $\sigma\in\mathcal{B%
}_{\tau}$. The maximal value $\ln\,\mathrm{rank\,}\mathcal{B}$ of the von
Neumann entropy is defined by the dimensionality $\mathrm{rank\,}\mathcal{B}%
=\dim\mathcal{A}^{0}$ of the maximal Abelian subalgebra of the decomposable
algebra $\mathcal{B}$, i.e. by $\dim\mathcal{H}$.

However, if $\pi $ is not c-coupling, the difference $\mathsf{S}\left( \pi
\right) =\mathsf{S}\left( \sigma \right) -\mathsf{I}\left( \pi \right) $ can
achieve the negative value, and may not serve as a measure of conditional
entropy in such a case.

\begin{definition}
The supremum of the mutual information 
\begin{equation}
\mathsf{H}\left( \varsigma \right) =\sup \left\{ \mathsf{I}\left( \pi
\right) :\mu \circ \pi =\varsigma \right\} =\mathsf{I}\left( \pi _{q}\right)
,  \label{4.8}
\end{equation}
which is achieved on $\mathcal{A}=\widetilde{\mathcal{B}}$ for a fixed state 
$\varsigma \left( B\right) =\mathrm{Tr}_{\mathcal{H}}B\sigma $ by the
standard q-coupling $\pi _{q}\left( B\right) =\sigma ^{1/2}\widetilde{B}%
\sigma ^{1/2}$, is called \emph{q-entropy} of the state $\varsigma $. The
maximum 
\begin{equation*}
\mathsf{S}\left( \varsigma \right) =\sup \left\{ \mathsf{I}\left( \pi
_{c}\right) :\mu \circ \pi _{c}=\varsigma \right\} =\mathsf{I}\left( \pi
_{d}^{0}\right)
\end{equation*}
over all c-couplings $\pi _{c}$ corresponding to c-encodings (\ref{2.5}),
which is achieved on an extreme d-coupling $\pi _{d}^{0}$, is called the 
\emph{c-entropy} of the state $\varsigma $. The differences, 
\begin{equation*}
\mathsf{H}\left( \pi \right) =\mathsf{H}\left( \varsigma \right) -\mathsf{I}%
\left( \pi \right) ,\quad \mathsf{S}\left( \pi \right) =\mathsf{S}\left(
\varsigma \right) -\mathsf{I}\left( \pi \right) ,
\end{equation*}
are called respectively, the \emph{q-conditional entropy} on $\mathcal{B}$
with respect to $\mathcal{A}$ and the \emph{(degree of) disentanglement} for
the coupling $\pi :\mathcal{B}\rightarrow \mathcal{A}$. A compound state is
said to be \emph{essentially entangled} if $\mathsf{S}\left( \pi \right) <0$%
, and $\mathsf{S}\left( \pi \right) \geq 0$ for a c-coupling $\pi =\pi _{c}$
(this is called the \emph{c-conditional entropy} on $\mathcal{B}$ with
respect to $\mathcal{A}$).
\end{definition}

Obviously, $\mathsf{H}\left( \varsigma \right) $ and $\mathsf{S}\left(
\varsigma \right) $ are both positive, do not depend, unlike $\mathsf{S}%
\left( \sigma \right) =\mathsf{S}\left( \varsigma /\nu \right) $, on the
choice of the faithful trace $\nu $ on $\mathcal{B}$ and obey the inequality 
$\mathsf{H}\left( \varsigma \right) \geq \mathsf{S}\left( \varsigma \right) $%
. The same is true for the conditional entropies $\mathsf{H}\left( \pi
\right) $ and $\mathsf{S}\left( \pi \right) $, where $\mathsf{S}\left( \pi
\right) $ always has a positive value 
\begin{equation*}
\mathsf{S}\left( \pi \right) \geq \mathsf{S}\left( \pi ^{0}\right) \geq 0
\end{equation*}
in the case of a c-coupling $\pi =\pi _{c}$ due to $\pi _{c}^{\ast }=\pi
_{d}^{\ast }\mathrm{K}$ for a normal unital CP map $\mathrm{K}:\mathcal{A}%
\rightarrow \mathcal{A}^{0}$, where $\pi ^{0}=\pi _{d}$ is a d-coupling with
Abelian $\mathcal{A}^{0}$. But the disentanglement $\mathsf{S}\left( \pi
\right) $ can also achieve the negative value 
\begin{equation}
\inf \left\{ \mathsf{S}\left( \pi \right) :\mu \circ \pi =\varsigma \right\}
=\mathsf{S}\left( \varsigma \right) -\mathsf{H}\left( \varsigma \right)
=-\sum_{i}\varkappa \left( i\right) \mathsf{S}\left( \sigma _{i}\right)
\label{4.9}
\end{equation}
as the following theorem states in the case of discrete $\mathcal{B}$. Here
the $\sigma _{i}\in \mathcal{L}\left( \mathcal{H}_{i}\right) $ are the
density operators of the normalized factor-states $\varsigma _{i}=\varkappa
\left( i\right) ^{-1}\varsigma |\mathcal{L}\left( \mathcal{H}_{i}\right) $
with $\varkappa \left( i\right) =\varsigma \left( I^{i}\right) $, where $%
I^{i}$ are the orthoprojectors onto $\mathcal{H}_{i}$. Note that $\mathsf{H}%
\left( \varsigma \right) =\mathsf{S}\left( \varsigma \right) $ if the
algebra $\mathcal{B}$ is completely decomposable, i.e. Abelian. In this case
the maximal value $\ln \,\mathrm{rank\,}\mathcal{B}$ of $\mathsf{S}\left(
\varsigma \right) $ can be written as $\ln \dim \mathcal{B}$. The
disentanglement $\mathsf{S}\left( \pi \right) $ is always positive in this
case, and $\mathsf{S}\left( \pi \right) =\mathsf{H}\left( \pi \right) $ as
in the case of Abelian $\mathcal{A}$.

\begin{theorem}
Let $\mathcal{B}$ be a discrete decomposable algebra on $\mathcal{H}=\oplus
_{i}\mathcal{H}_{i}$, with a normal state given by the density operator $%
\sigma =\oplus \sigma \left( i\right) $ with respect to the trace $\mu =%
\mathrm{Tr}_{\mathcal{H}}$ on $\mathcal{B}$, and $\mathcal{C}\subseteq 
\mathcal{B}$ be its center with the state $\varkappa =\varsigma |\mathcal{C}$
induced by the probability distribution $\varkappa \left( i\right) =\mathrm{%
Tr\,}\sigma \left( i\right) .$ Then the c-entropy $\mathsf{S}\left(
\varsigma \right) $ is given as the von Neumann entropy (\ref{4.6}) of the
density operator $\sigma $ and the q-entropy (\ref{4.8}) is given by the
formula 
\begin{equation}
\mathsf{H}\left( \varsigma \right) =\sum_{i}\left( \varkappa \left( i\right)
\ln \varkappa \left( i\right) -2\mathrm{Tr}_{\mathcal{H}_{i}}\sigma \left(
i\right) \ln \sigma \left( i\right) \right) .  \label{4.10}
\end{equation}
This can be written as $\mathsf{H}\left( \varsigma \right) =\mathsf{H}_{%
\mathcal{B}|\mathcal{C}}\left( \varsigma \right) +\mathsf{H}_{\mathcal{C}%
}\left( \varsigma \right) $, where $\mathsf{H}_{\mathcal{C}}\left( \varsigma
\right) =-\sum_{i}\varkappa \left( i\right) \ln \varkappa \left( i\right) $,
and 
\begin{equation*}
\mathsf{H}_{\mathcal{B}|\mathcal{C}}\left( \varsigma \right)
=-2\sum_{i}\varkappa \left( i\right) \mathrm{Tr}_{\mathcal{H}_{i}}\sigma
_{i}\ln \sigma _{i}=2\mathsf{S}_{\mathcal{B}|\mathcal{C}}\left( \varsigma
\right) ,
\end{equation*}
with $\sigma _{i}=\sigma \left( i\right) /\varkappa \left( i\right) $. $%
\mathsf{H}\left( \varsigma \right) $ is finite iff $\mathsf{S}\left(
\varsigma \right) <\infty $, and if $\mathcal{B}$ is finite-dimensional, it
is bounded, with maximal value $\mathsf{H}\left( \varsigma ^{\circ }\right)
=\ln \dim \mathcal{B}$, achieved for $\sigma ^{\circ }=\oplus \sigma
_{i}^{\circ }\varkappa ^{\circ }\left( i\right) $ 
\begin{equation*}
\sigma _{i}^{\circ }=\left( \dim \mathcal{H}_{i}\right) ^{-1}I^{i},\quad
\varkappa ^{\circ }\left( i\right) =\dim \mathcal{B}\left( i\right) /\dim 
\mathcal{B},
\end{equation*}
where $\dim \mathcal{B}\left( i\right) =\left( \dim \mathcal{H}_{i}\right)
^{2}$, $\dim \mathcal{B}=\sum_{i}\dim \mathcal{B}\left( i\right) $.
\end{theorem}

\proof%
%
We have already proven that $\mathsf{S}\left( \varsigma\right) =\mathsf{S}%
\left( \sigma\right) $, where 
\begin{equation*}
\mathsf{S}\left( \sigma\right) =-\sum_{i}\mathrm{Tr}_{\mathcal{H}%
_{i}}\sigma\left( i\right) \ln\sigma\left( i\right) =\mathsf{S}_{\mathcal{C}%
}\left( \varsigma\right) +\mathsf{S}_{\mathcal{B}|\mathcal{C}}\left(
\varsigma\right) ,
\end{equation*}
with $\mathsf{S}_{\mathcal{C}}\left( \varsigma\right) =\mathsf{H}_{\mathcal{C%
}}\left( \varsigma\right) $, $\mathsf{S}_{\mathcal{B}|\mathcal{C}}\left(
\varsigma\right) =\sum\varkappa\left( i\right) \mathsf{S}\left(
\sigma_{i}\right) =\frac{1}{2}\mathsf{H}_{\mathcal{B}|\mathcal{C}}\left(
\varsigma\right) $.

The q-entropy $\mathsf{H}\left( \varsigma\right) $ is the supremum (\ref{4.8}%
) of the mutual information (\ref{4.3}) which is achieved on the standard
entanglement, corresponding to the density operator $\omega
=\oplus\omega\left( i,k\right) $ with $\omega\left( i,k\right)
=\varkappa\left( i\right) |\sigma_{i}^{1/2})(\sigma_{i}^{1/2}|\delta_{k}^{i}$
of the standard compound state (\ref{1.10}) with $\widetilde {\mathcal{B}}=%
\mathcal{B}$, $\rho=\sigma$. Thus $\mathsf{H}\left( \varsigma\right) =%
\mathsf{I}\left( \pi_{q}\right) $, where 
\begin{align*}
\mathsf{I}\left( \pi_{q}\right) & =\mathrm{Tr}\omega\left( \ln\omega
-\ln\left( \sigma\otimes I\right) -\ln\left( I\otimes\sigma\right) \right) =%
\mathsf{S}\left( \omega\right) -2\mathsf{S}\left( \sigma\right) \\
& =\sum_{i}\varkappa\left( i\right) \ln\varkappa\left( i\right) -2\mathrm{%
Tr\,}\sigma\ln\sigma=-\sum_{i}\varkappa\left( i\right) \left(
\ln\varkappa\left( i\right) +2\mathrm{Tr}_{\mathcal{H}_{i}}\sigma_{i}\ln%
\sigma_{i}\right) .
\end{align*}
Here we used that $\mathrm{Tr\,\,}\omega\ln\omega=\sum_{i}\varkappa\left(
i\right) \ln\varkappa\left( i\right) $ due to 
\begin{equation*}
\omega\ln\omega=\oplus_{i,k}\omega\left( i,k\right) \ln\omega\left(
i,k\right) =\oplus_{i}\varkappa\left( i\right)
|\sigma_{i}^{1/2})(\sigma_{i}^{1/2}|\ln\varkappa\left( i\right) ,
\end{equation*}
and that $\mathrm{Tr\,\,}\sigma\ln\sigma=\sum_{i}\varkappa\left( i\right)
\left( \ln\varkappa\left( i\right) -\mathsf{S}_{\mathcal{B}_{i}}\left(
\varsigma_{i}\right) \right) $ due to 
\begin{equation*}
\sigma\ln\sigma=\oplus_{i}\sigma\left( i\right) \ln\sigma\left( i\right)
=\oplus_{i}\varkappa\left( i\right) \sigma_{i}\left( \ln\varkappa\left(
i\right) +\ln\sigma_{i}\right)
\end{equation*}
for the orthogonal decomposition $\sigma=\oplus_{i}\varkappa\left( i\right)
\sigma_{i}$, where $\varkappa\left( i\right) =\mathrm{Tr\,}\sigma\left(
i\right) $.

Thus $\mathsf{H}\left( \varsigma\right) =\mathsf{H}_{\mathcal{B}|\mathcal{C}%
}\left( \varsigma\right) +\mathsf{H}_{\mathcal{C}}\left( \varsigma\right) =2%
\mathsf{S}_{\mathcal{B}|\mathcal{C}}\left( \varsigma \right) +\mathsf{S}_{%
\mathcal{C}}\left( \varsigma\right) \leq 2\mathsf{S}\left( \varsigma\right) $%
, and it is bounded by 
\begin{align*}
\mathsf{C}_{\mathcal{B}} & =\sup_{\varkappa}\sum_{i}\varkappa\left( i\right)
\left( 2\sup_{\varsigma_{i}}\mathsf{S}_{\mathcal{B}\left( i\right) }\left(
\varsigma_{i}\right) -\ln\varkappa\left( i\right) \right) \\
& =-\inf_{\varkappa}\sum_{i}\varkappa\left( i\right) \left( \ln
\varkappa\left( i\right) -2\ln\mathrm{\dim}\mathcal{H}_{i}\right) =\ln \dim%
\mathcal{B}.
\end{align*}
Here we used the fact that the supremum of von Neumann entropies $\mathsf{S}%
\left( \sigma_{i}\right) $ for the simple algebras $\mathcal{B}\left(
i\right) =\mathcal{L}\left( \mathcal{H}_{i}\right) $ with $\dim\mathcal{B}%
\left( i\right) =\left( \dim\mathcal{H}_{i}\right) ^{2}<\infty$ is achieved
on the tracial density operators $\sigma_{i}=\left( \dim\mathcal{H}%
_{i}\right) ^{-1}I^{i}\equiv\sigma_{i}^{\circ}$, and the infimum of the
relative entropy 
\begin{equation*}
\mathsf{R}\left( \varkappa:\varkappa^{\circ}\right) =\sum_{i}\varkappa
\left( i\right) \left( \ln\varkappa\left( i\right) -\ln\varkappa^{\circ
}\left( i\right) \right) ,
\end{equation*}
where $\varkappa^{\circ}\left( i\right) =\dim\mathcal{B}\left( i\right) /\dim%
\mathcal{B}$, is zero, achieved at $\varkappa=\varkappa^{\circ}$.%
\endproof%
%

Note that as shown in \cite{OP93} for the case of the simple algebra $%
\mathcal{B}$, the quantum enropy $\mathsf{H}\left( \varsigma\right) $ can be
also achieved as the supremum of the von Neumann entropy $\mathsf{S}\left(
\varrho\right) $ over all pure couplings given by the isometries $X:\mathcal{%
H}\rightarrow\mathcal{G}\otimes\mathcal{H}$, $X^{\dagger}X=I$ preserving the
state $\varsigma$. The latter means that the density operator $\omega$ of
the corresponding compound states with the marginals $\rho=\mathrm{Tr}_{%
\mathcal{H}}\omega$ and $\sigma=\mathrm{Tr}_{\mathcal{G}}\omega$ is given as 
$\omega=X\sigma X^{\dagger}$.

\section{Quantum Channel and Entropic Capacities}

In this section we describe quantum noisy channel in terms of normal unital
CP maps and their duals, and introduce an analog of Shannon information for
general semifinite algebras. We consider the maximization problems for this
quantity with various operational constrains on encodings, and define the
entropic capacities which serve as upper bounds for the operational
capacities corresponding to these constrains. The question of asymptotic
equivalence of the entropic and operational capacities is not touched here.

Let $\mathcal{H}_{1}$ be a Hilbert space describing a quantum input system
and $\mathcal{H}$ describe its output Hilbert space. A quantum channel is an
affine operation sending each input state defined on $\mathcal{H}_{1}$ to an
output state defined on $\mathcal{H}$ such that the mixtures of states are
preserved. A deterministic quantum channel is given by a linear isometry $U%
\mathrm{:\mathcal{H}}_{1}\rightarrow\mathrm{\mathcal{H}}$ with $U^{\dagger
}U=I^{1}$ ($I^{1}$ is the identify operator in $\mathrm{\mathcal{H}}_{1}$)
such that each input state vector $\eta_{1}\in\mathrm{\mathcal{H}}_{1}$, $%
\left\| \eta_{1}\right\| =1$, is transmitted into an output state vector $%
\eta=U\eta_{1}\in\mathcal{H}$, $\left\| \eta\right\| =1$. The orthogonal
sums $\varsigma_{1}=\oplus\varsigma_{1}\left( n\right) $ of pure input
states $\varsigma_{1}\left( B,n\right) =\eta_{1}\left( n\right) ^{\dagger
}B\eta_{1}\left( n\right) $ are sent into the orthogonal sums $\varsigma
=\oplus\varsigma\left( n\right) $ of pure states on $\mathcal{B}=\mathcal{L}%
\left( \mathcal{H}\right) $ corresponding to the orthogonal state vectors $%
\eta\left( n\right) =U\eta_{1}\left( n\right) $.

A noisy quantum channel sends pure input states $\varsigma_{1}$ on an
algebra $\mathcal{B}^{1}\subseteq\mathcal{L}\left( \mathcal{H}_{1}\right) $
into mixed ones $\varsigma=\varsigma_{1}\Lambda$ given by the composition
with a normal completely positive unital map $\Lambda:\mathcal{B}\rightarrow 
\mathcal{B}^{1}$. We shall assume that $\mathcal{B}^{1}$ (as well as $%
\mathcal{B}$) is equipped with a normal faithful semifinite trace $\nu_{1}$
defining the pairing $\left\langle B,u^{\dagger}u\right\rangle _{1}=\nu
_{1}\left( \tilde{u}^{\dagger}B\tilde{u}\right) $ of $\mathcal{B}^{1}$ and $%
\mathcal{B}_{\intercal}^{1}=\widetilde{\mathcal{B}_{\ast}^{1}}$. Then the
input-output state transformations are described by the transposed map $%
\Lambda^{\intercal}:\mathcal{B}_{\intercal}^{1}\rightarrow\mathcal{B}%
_{\intercal}$%
\begin{equation*}
\left\langle B,\Lambda^{\intercal}\left( \sigma_{1}\right) \right\rangle
=\left\langle \Lambda\left( B\right) ,\sigma_{1}\right\rangle _{1},\quad B\in%
\mathcal{B},\sigma_{1}\in\mathcal{B}_{\intercal}^{1}
\end{equation*}
defining the output density operators $\sigma=\Lambda^{\intercal}\left(
\sigma_{1}\right) $ for any input normal state $\varsigma_{1}\left( B\right)
=\left\langle B,\sigma_{1}\right\rangle _{1}$. Without loss of generality
the input algebra $\mathcal{B}^{1}$ can be assumed to be the smallest
decomposable algebra generated by the range $\Lambda\left( \mathcal{B}%
\right) $ of the channel map $\Lambda$ ($\mathcal{B}^{1}$ is Abelian if $%
\Lambda\left( \mathcal{B}\right) $ consists of only commuting operators on $%
\mathcal{H}_{1}$).

The input generalized entanglements $\varpi^{1}:\mathcal{A}\rightarrow 
\mathcal{B}_{\intercal}^{1}$, including encodings of the state $%
\varsigma_{1} $ with the density $\sigma_{1}=\varpi^{1}\left( I\right) $,
will be defined by the couplings $\kappa^{\ast}:\mathcal{B}^{1}\rightarrow%
\mathcal{A}_{\ast}$ as $\varpi^{1}=\kappa^{\symbol{126}}$. Here $\kappa:%
\mathcal{A}\rightarrow \mathcal{B}_{\ast}^{1}$ is a normal TCP map defining
the state $\varrho =\nu_{1}\circ\kappa$ of a probe system $\left( \mathcal{A}%
,\mu\right) $ which is entangled to $\left( \mathcal{B}^{1},\varsigma_{1}%
\right) $ by $\kappa^{\symbol{126}}\left( A\right) =J\kappa\left(
A^{\dagger}\right) J$, and the adjoint map $\kappa^{\ast}$ is defined as
usual by 
\begin{equation*}
\left\langle A|\kappa^{\ast}\left( B\right) \right\rangle _{\mu}=\omega
_{1}\left( A^{\dagger}\otimes B\right) =\left\langle \kappa\left( A\right)
|B\right\rangle _{1},\quad\forall A\in\mathcal{A},B\in\mathcal{B}_{1},
\end{equation*}
where $\omega_{1}$ is the corresponding compound state on $\mathcal{A}\otimes%
\mathcal{B}^{1}$.

These (generalized) entanglements describe the quantum-quantum
correspondences (q-, c-, or o-encodings) of the probe systems $\left( 
\mathcal{A},\varrho\right) $ with the density operators $\rho=\kappa^{%
\intercal}\left( I^{1}\right) $, to the input $\left( \mathcal{B}%
^{1},\varsigma_{1}\right) $ of the channel $\Lambda$. In particular, the
most informative standard input entanglement $\varpi_{q}^{1}:\widetilde{%
\mathcal{B}^{1}}\rightarrow \mathcal{B}_{\intercal}^{1}$ is the entanglement
of the transposed input system $\left( \mathcal{A}^{0},\varrho_{0}\right)
=\left( \widetilde {\mathcal{B}^{1}},\widetilde{\varsigma_{1}}\right) $
corresponding to the TCP map $\kappa_{q}\left( A\right)
=J\sigma_{1}^{1/2}A^{\dagger}\sigma_{1}^{1/2}J$. In the case of discrete
decomposable $\mathcal{A}^{0}=\widetilde {\mathcal{B}^{1}}=\mathcal{B}^{1}$
with the density operator $\sigma _{1}=\oplus_{i}\sigma_{1}\left( i\right) $
this extreme input q-encoding defines the following density operator 
\begin{equation}
\omega_{q}=\left( \mathrm{I}\otimes\Lambda^{\intercal}\right) \left(
\omega_{q1}\right) ,\quad\omega_{q1}=\oplus_{i}|\sigma_{1}\left( i\right)
^{1/2})(\sigma_{1}\left( i\right) ^{1/2}|  \label{3.1}
\end{equation}
of the input-output compound state $\omega_{q1}\Lambda$ on $\mathcal{A}%
^{0}\otimes\mathcal{B}=\mathcal{B}^{1}\otimes\mathcal{B}$.

The other extreme case of the generalized input entanglements, the pure
c-encodings corresponding to (\ref{2.2}), are less informative then the pure
d-encodings $\varpi_{d}^{1}=\kappa_{d}^{\symbol{126}}$ given by the
decompositions $\kappa_{d}^{\ast}=\sum|n\rangle\langle n|\varsigma_{1}\left(
n\right) $ with pure states $\varsigma_{1}\left( B,n\right) =\eta\left(
n\right) ^{\dagger}B\eta\left( n\right) $ on $\mathcal{B}_{1}$. They define
the density operators 
\begin{equation}
\omega_{d}=\left( \mathrm{I}\otimes\Lambda^{\intercal}\right) \left(
\omega_{d1}\right) ,\quad\omega_{d1}=\sum_{n}|n\rangle\langle n|\otimes
\eta_{1}\left( n\right) \eta_{1}\left( n\right) ^{\dagger},  \label{3.2}
\end{equation}
of the $\mathcal{B}^{1}\otimes\mathcal{B}$-compound state $%
\omega_{d1}\Lambda=\omega_{d1}\circ\left( \mathrm{I}\otimes\Lambda\right) $.
These are the Ohya compound states $\omega_{o}=\omega_{o1}\Lambda$ \cite
{Ohy83} in the case 
\begin{equation*}
\sigma_{1}\left( n\right) =\eta_{1}^{o}\left( n\right) \eta_{1}^{o}\left(
n\right) ^{\dagger},\quad\eta_{1}^{o}\left( n\right)
^{\dagger}\eta_{1}^{o}\left( m\right) =p_{1}\left( n\right) \delta_{n}^{m},
\end{equation*}
of orthogonality of the density operators $\sigma_{1}\left( n\right) $
normalized to the eigen-values $p_{1}\left( n\right) $ of $\sigma_{1}$. The
o-compound states are achieved by pure o-encodings $\varpi_{d}^{1}=\kappa
_{o}^{\symbol{126}}$ described by the couplings $\kappa_{o}=\sum
|n\rangle\langle n|\varsigma_{1}^{o}\left( n\right) $ with $\varsigma
_{1}^{o}$\ corresponding to $\eta_{1}^{o}$. The\ input-output density
operator 
\begin{equation}
\omega_{o}=\left( \mathrm{I}\otimes\Lambda^{\intercal}\right) \omega
_{o1},\quad\omega_{o1}=\sum_{n}|n\rangle\langle n|\otimes\eta_{1}^{o}\left(
n\right) \eta_{1}^{o}\left( n\right) ^{\dagger}  \label{3.3}
\end{equation}
of the Ohya compound state $\omega_{o}$ is achieved by the coupling $%
\lambda=\kappa^{\ast}\Lambda$ of the output $\left( \mathcal{B}%
,\varsigma\right) $ to the extreme probe system $\left( \mathcal{A}%
^{0},\varrho_{0}\right) =\left( \mathcal{B}^{1},\varsigma_{1}\right) $ as
the composition of $\kappa^{\ast}$ and the channel $\Lambda$.

If $\mathrm{K}:\mathcal{A}\rightarrow\mathcal{A}^{0}$ is a normal completely
positive unital map 
\begin{equation*}
\mathrm{K}\left( A\right) =\mathrm{Tr}_{\mathcal{F}_{-}}\widetilde {X}A%
\widetilde{X}^{\dagger},\quad A\in\mathcal{A},
\end{equation*}
where $X$ is a bounded operator $\mathcal{F}_{-}\otimes\mathcal{G}%
_{0}\rightarrow\mathcal{G}$ with $\mathrm{Tr}_{\mathcal{F}_{-}}X^{\dagger
}X=I^{0}$, the compositions $\kappa=\kappa_{0}\mathrm{K}$, $%
\pi=\Lambda^{\ast }\kappa$ describe the entanglements of the probe system $%
\left( \mathcal{A},\varrho\right) $ to the channel input $\left( \mathcal{B}%
^{1},\varsigma_{1}\right) $ and the output $\left( \mathcal{B}%
,\varsigma\right) $ via this channel respectively. The state $\varrho
=\varrho_{0}\mathrm{K}$ is given by 
\begin{equation*}
\mathrm{K}^{\intercal}\left( \rho_{0}\right) =X\left( I^{-}\otimes\rho
_{0}\right) X^{\dagger}\in\mathcal{A}_{\ast}
\end{equation*}
for each density operator $\rho_{0}\in\mathcal{A}_{\ast}^{0}$, where $I^{-}$
is the identity operator in $\mathcal{F}_{-}$. The resulting entanglement $%
\pi=\lambda^{\ast}\mathrm{K}$ defines the compound state $\omega=\omega
_{01}\circ\left( \mathrm{K}\otimes\Lambda\right) $ on $\mathcal{A}\otimes%
\mathcal{B}$ with 
\begin{equation*}
\omega_{01}\left( A^{0}\otimes B^{1}\right) =\mathrm{Tr\,\,}\tilde{A}%
^{0}\kappa_{0}^{\ast}\left( B^{1}\right) =\mathrm{Tr\,\,}\tilde{\upsilon }%
_{01}^{\dagger}\left( A^{0}\otimes B^{1}\right) \tilde{\upsilon}_{01}
\end{equation*}
on $\mathcal{A}^{0}\otimes\mathcal{B}^{1}$. Here $\upsilon_{01}:\mathcal{G}%
_{0}\otimes\mathrm{\mathcal{H}}_{1}\rightarrow\mathcal{F}_{01}$ is the
amplitude operator uniquely defined by the input compound density operator $%
\omega_{01}\in\mathcal{A}_{\intercal}^{0}\otimes\mathcal{B}_{\intercal}^{1}$
up to a unitary operator $U^{0}$ on $\mathcal{F}_{01}$. The effect of the
input entanglement $\kappa$ and the output channel $\Lambda$ can be written
in terms of the amplitude operator of the state $\omega$ as 
\begin{equation*}
\upsilon=\left( X\otimes Y\right) \left( I^{-}\otimes\upsilon_{01}\otimes
I^{+}\right) U
\end{equation*}
up to a unitary operator $U$ in $\mathcal{F}=\mathcal{F}_{-}\otimes \mathcal{%
F}_{01}\otimes\mathcal{F}_{+}$. Thus the density operator of the
input-output compound state $\omega$ is given by $\omega_{01}\left( \mathrm{K%
}\otimes\Lambda\right) $ with the density 
\begin{equation}
\left( \mathrm{K}\otimes\Lambda\right) ^{\ast}\left( \omega_{01}\right)
=\left( X\otimes Y\right) \omega_{01}\left( X\otimes Y\right) ^{\dagger },
\label{3.4}
\end{equation}
where $\omega_{01}=\upsilon_{01}\upsilon_{01}^{\dagger}$.

Let $\mathcal{K}_{q}^{1}$ be the set of all normal TCP maps $\kappa :%
\mathcal{A}\rightarrow\mathcal{B}_{\ast}^{1}$ with any probe algebra $%
\mathcal{A}$ normalized as $\mathrm{Tr}\kappa\left( I\right) =1$ and $%
\mathcal{K}_{q}\left( \varsigma_{1}\right) $ be the subset of all$\ \kappa\in%
\mathcal{K}_{q}^{1}$ with $\kappa\left( I\right) =\varsigma_{1}$. Each $%
\kappa\in\mathcal{K}_{q}^{1}$ can be decomposed as $\kappa_{q}\mathrm{K}$,
where $\kappa_{q}:\mathcal{A}^{0}\rightarrow \mathcal{B}^{1}$ defines the
standard input entanglement $\varpi_{q}^{1}=\kappa_{q}^{\symbol{126}}$, and $%
\mathrm{K}$ is a normal unital CP map $\mathcal{A}\rightarrow\widetilde{%
\mathcal{B}^{1}}$.

Further let $\mathcal{K}_{c}^{1}$ be the set of all CP-TCP maps $\kappa$
described as the combinations 
\begin{equation}
\kappa\left( A\right) =\sum_{n}\varrho_{n}\left( A\right) \sigma _{1}\left(
n\right)  \label{3.6}
\end{equation}
of the primitive maps $A\mapsto\varrho_{n}\left( A\right) \sigma_{1}\left(
n\right) $, and $\mathcal{K}_{d}^{1}$ be the subset of the diagonalizing
entanglements $\kappa$, i.e. the decompositions 
\begin{equation}
\kappa\left( A\right) =\sum_{n}\langle n|A|n\rangle\sigma_{1}\left( n\right)
.  \label{3.5}
\end{equation}
As in the first case $\mathcal{K}_{c}\left( \varsigma_{1}\right) $ and $%
\mathcal{K}_{d}\left( \varsigma_{1}\right) $ denote the subsets
corresponding to a fixed $\kappa\left( I\right) =\varsigma_{1}$. Each $%
\mathcal{K}_{c}\left( \varsigma_{1}\right) $ can be represented as the
composition $\kappa=\kappa_{d}\mathrm{K}$, where $\kappa_{d}$ normalized to $%
\varsigma_{1}$ describes a pure d-encoding $\varpi_{d}^{1}=\kappa _{d}^{%
\symbol{126}}$ of $\left( \mathcal{B}^{1},\varsigma_{1}\right) $ for a
proper choice of the CP map $\mathrm{K}:\mathcal{A}\rightarrow\mathcal{B}%
^{1} $.

Furthermore let $\mathcal{K}_{o}^{1}$ (and $\mathcal{K}_{o}\left(
\varsigma_{1}\right) $) be the subset of all decompositions (\ref{3.5}) with
orthogonal $\sigma_{1}\left( n\right) $ (and fixed $\sum_{n}\sigma
_{1}\left( n\right) =\sigma_{1}$): 
\begin{equation*}
\quad\sigma_{1}\left( m\right) \sigma_{1}\left( n\right) =0,\,m\neq n.
\end{equation*}
Each $\kappa\in\mathcal{K}_{o}\left( \varsigma_{1}\right) $ can also be
represented as $\kappa=\kappa_{o}\mathrm{K}$, with $\kappa_{o}$ describing
the pure o-encoding $\varpi_{o}^{1}=\kappa_{o}^{\symbol{126}}$ of $\left( 
\mathcal{B}^{1},\varsigma_{1}\right) =\left( \mathcal{A}^{0},\varrho
_{0}\right) $.

Now, let us maximize the entangled mutual entropy for a given quantum
channel $\Lambda $ (and a fixed input state $\varsigma _{1}$ on the
decomposable $\mathcal{B}^{1}=\widetilde{\mathcal{B}^{1}}$) by means of the
above four types of entanglement $\kappa $. The mutual information (\ref{4.3}%
) was defined in the previous section by the density operators of the
corresponding compound state $\omega $ on $\mathcal{A}\otimes \mathcal{B}$
and the product-state $\varphi =\varrho \otimes \varsigma $ of the marginals 
$\varrho $, $\varsigma $ for $\omega $. In each case 
\begin{equation*}
\omega =\omega _{01}\left( \mathrm{K}\otimes \Lambda \right) ,\quad \varphi
=\varphi _{01}\left( \mathrm{K}\otimes \Lambda \right) ,
\end{equation*}
where $\mathrm{K}$ is a CP map $\mathcal{A}\rightarrow \mathcal{A}^{0}=%
\mathcal{B}^{1}$, $\omega _{01}$ is one of the corresponding extreme
compound states $\omega _{q1}$, $\omega _{c1}=\omega _{d1}$, $\omega _{o1}$
on $\mathcal{B}^{1}\otimes \mathcal{B}^{1}$, and $\varphi _{01}=\rho
_{0}\otimes \varsigma _{1}$. The density operator $\omega =\left( \mathrm{K}%
\otimes \Lambda \right) ^{\intercal }\left( \omega _{01}\right) $ is written
in (\ref{3.4}), and $\phi =\rho \otimes \sigma $ can be written as 
\begin{equation*}
\phi =\kappa ^{\intercal }\left( I\right) \otimes \lambda ^{\intercal
}\left( I\right) ,
\end{equation*}
where $\lambda ^{\intercal }=\Lambda ^{\intercal }\pi _{1}^{0}$. This proves
the following proposition.

\begin{proposition}
The entangled mutual informations achieve the following maximal values 
\begin{equation}
\sup_{\kappa \in \mathcal{K}_{q}\left( \varsigma _{1}\right) }\mathsf{I}%
\left( \kappa ^{\ast }\Lambda \right) =\mathsf{I}_{q}\left( \varsigma
_{1},\Lambda \right) :=\mathsf{I}\left( \kappa _{q}^{\ast }\Lambda \right) ,
\label{3.7}
\end{equation}
\begin{equation*}
\mathsf{I}_{c}\left( \varsigma _{1},\Lambda \right) :=\sup_{\kappa \in 
\mathcal{K}_{c}\left( \varsigma _{1}\right) }\mathsf{I}\left( \kappa ^{\ast
}\Lambda \right) =\sup_{\kappa _{d}}\mathsf{I}\left( \kappa _{d}^{\ast
}\Lambda \right) \equiv \mathsf{I}_{d}\left( \varsigma _{1},\Lambda \right) ,
\end{equation*}
\begin{equation}
\sup_{\kappa \in \mathcal{K}_{o}\left( \varsigma _{1}\right) }\mathsf{I}%
\left( \kappa ^{\ast }\Lambda \right) =\mathsf{I}_{o}\left( \varsigma
_{1},\Lambda \right) :=\sup_{\kappa _{o}}\mathsf{I}\left( \kappa _{o}^{\ast
}\Lambda \right) ,  \label{3.8}
\end{equation}
where $\kappa _{\cdot }$ are the corresponding extremal input couplings $%
\mathcal{A}^{0}\rightarrow \mathcal{B}_{\ast }^{1}$ with $\mu \circ \kappa
_{\cdot }^{\ast }=\varsigma _{1}$. They are ordered as 
\begin{equation}
\mathsf{I}_{q}\left( \varsigma _{1},\Lambda \right) \geq \mathsf{I}%
_{c}\left( \varsigma _{1},\Lambda \right) =\mathsf{I}_{d}\left( \varsigma
_{1},\Lambda \right) \geq \mathsf{I}_{o}\left( \varsigma _{1},\Lambda
\right) .  \label{3.9}
\end{equation}
\end{proposition}

In the following definition the maximal information $\mathsf{I}_{c}\left(
\varsigma _{1},\Lambda \right) =\mathsf{I}_{d}\left( \varsigma _{1},\Lambda
\right) $ is simply denoted as $\mathsf{I}_{1}\left( \varsigma _{1},\Lambda
\right) $.

\begin{definition}
The suprema 
\begin{equation*}
\mathsf{C}_{q}\left( \Lambda \right) =\sup_{\kappa \in \mathcal{K}_{q}^{1}}%
\mathsf{I}\left( \kappa ^{\ast }\Lambda \right) =\sup_{\varsigma _{1}}%
\mathsf{I}_{q}\left( \varsigma _{1},\Lambda \right) ,\;
\end{equation*}
\begin{equation}
\sup_{\kappa \in \mathcal{K}_{d}^{1}}\mathsf{I}\left( \kappa ^{\ast }\Lambda
\right) =\mathsf{C}_{1}\left( \Lambda \right) :=\sup_{\varsigma _{1}}\mathsf{%
I}_{1}\left( \varsigma _{1},\Lambda \right) ,\;  \label{3.10}
\end{equation}
\begin{equation*}
\mathsf{C}_{o}\left( \Lambda \right) =\sup_{\kappa \in \mathcal{K}_{o}^{1}}%
\mathsf{I}\left( \kappa ^{\ast }\Lambda \right) =\sup_{\varsigma _{1}}%
\mathsf{I}_{o}\left( \varsigma _{1},\Lambda \right) ,\;
\end{equation*}
are called the q-, c- or d-, and o-capacities respectively for the quantum
channel defined by a normal unital CP map $\Lambda :\mathcal{B}\rightarrow 
\mathcal{B}^{1}$.
\end{definition}

Obviously, the capacities (\ref{3.10}) satisfy the inequalities 
\begin{equation*}
\mathsf{C}_{o}\left( \Lambda \right) \leq \mathsf{C}_{1}\left( \Lambda
\right) \leq \mathsf{C}_{q}\left( \Lambda \right) .
\end{equation*}

\begin{theorem}
Let $\Lambda \left( B\right) =U^{\dagger }BU$ be a unital CP map $\mathcal{B}%
\rightarrow \mathcal{B}^{1}$ describing a quantum deterministic channel.
Then 
\begin{equation*}
\mathsf{I}_{1}\left( \varsigma _{1},\Lambda \right) =\mathsf{I}_{o}\left(
\varsigma _{1},\Lambda \right) =\mathsf{S}\left( \varsigma _{1}\right)
,\quad \mathsf{I}_{q}\left( \varsigma _{1},\Lambda \right) =\mathsf{S}%
_{q}\left( \varsigma _{1}\right) ,
\end{equation*}
where $\mathsf{S}_{q}\left( \varsigma _{1}\right) =\mathsf{H}\left(
\varsigma _{1}\right) $, and thus in this case 
\begin{equation*}
\mathsf{C}_{1}\left( \Lambda \right) =\mathsf{C}_{o}\left( \Lambda \right)
=\ln \mathrm{rank\,}\mathcal{B}^{1},\quad \mathsf{C}_{q}\left( \Lambda
\right) =\ln \dim \mathcal{B}^{1}.
\end{equation*}
\end{theorem}

\proof%
%
It was proved in the previous section for the case of the identity channel $%
\Lambda=\mathrm{I}$ and is thus also valid for any isomorphism $\Lambda
:B\mapsto U^{\dagger}BU$ describing the state transformations $\Lambda
^{\intercal}:\sigma\mapsto Y\sigma Y^{\dagger}$ by a unitary operator $U=%
\overline{Y}$. In the case of non-unitary $Y$ we can use the identity 
\begin{equation*}
\mathrm{Tr\,\,}Y\left( \sigma_{1}\otimes I^{+}\right) Y^{\dagger}\ln Y\left(
\sigma_{1}\otimes I^{+}\right) Y^{\dagger}=\mathrm{Tr\,\,}S\left(
\sigma_{1}\otimes I^{+}\right) \ln S\left( \sigma_{1}\otimes I^{+}\right) ,
\end{equation*}
where $S=Y^{\dagger}Y$. Due to this $\mathsf{S}\left(
\varsigma_{1}\Lambda\right) =-\mathrm{Tr\,\,}S\left( \sigma_{1}\otimes
I^{+}\right) \ln S\left( \sigma_{1}\otimes I^{+}\right) $, and $\mathsf{S}%
\left( \omega _{01}\left( \mathrm{K}\otimes\Lambda\right) \right) =$ 
\begin{equation*}
-\mathrm{Tr\,\,}\left( R\otimes S\right) \left( I^{-}\otimes\omega
_{01}\otimes I^{+}\right) \ln\left( R\otimes S\right) \left(
I^{-}\otimes\omega_{01}\otimes I^{+}\right) ,
\end{equation*}
where $R=X^{\dagger}X$. Thus $\mathsf{S}\left( \varsigma_{1}\Lambda\right) =%
\mathsf{S}\left( \varsigma_{1}\right) $, $\mathsf{S}\left( \omega
_{01}\left( \mathrm{K}\otimes\Lambda\right) \right) =\mathsf{S}\left(
\omega_{01}\left( \mathrm{K}\otimes\mathrm{I}\right) \right) $ if $%
Y^{\dagger}Y=I$, and 
\begin{align*}
\mathsf{I}\left( \left( \pi_{1}\Lambda\right) \right) & =\mathsf{S}\left(
\varrho_{0}\mathrm{K}\right) +\mathsf{S}\left( \varsigma_{1}\right) -\mathsf{%
S}\left( \omega_{01}\left( \mathrm{K}\otimes\mathrm{I}\right) \right) \\
& \leq\mathsf{S}\left( \varrho_{0}\right) +\mathsf{S}\left( \varsigma
_{1}\right) -\mathsf{S}\left( \omega_{01}\right) =\mathsf{I}\left(
\omega_{01}\right)
\end{align*}
for $\kappa=\kappa_{0}\mathrm{K}$ with any normal unital CP map $\mathrm{K}:%
\mathcal{A}\rightarrow\mathcal{A}^{0}$ and a compound state $\omega_{01}$ on 
$\mathcal{A}^{0}\otimes\mathcal{B}^{1}$. The supremum (\ref{3.7}), which is
achieved at the standard entanglement, corresponding to $\omega_{01}=%
\omega_{q1}$, coincides with q-entropy $\mathsf{H}\left( \varsigma
_{1}\right) $ and the supremum (\ref{3.8}), coinciding with $\mathsf{S}%
\left( \varsigma_{1}\right) $, is achieved for a pure o-entanglement,
corresponding to $\omega_{01}=\omega_{o1}$ given by any Schatten
decomposition for $\sigma_{1}$. Moreover, the entropy $\mathsf{H}\left(
\varsigma _{1}\right) $ is also achieved by any pure d-entanglement,
corresponding to $\omega_{01}=\omega_{d1}$ given by any extreme
decomposition for $\sigma_{1}$ and thus is the maximal mutual information $%
\mathsf{I}_{1}\left( \varsigma_{1},\Lambda\right) $ in the case of
deterministic $\Lambda$. Thus the capacity $\mathsf{C}_{1}\left(
\Lambda\right) $ of the deterministic channel is given by the maximum $%
\mathsf{C}_{o}=\ln\dim\mathcal{H}_{1}$ of the von Neumann entropy $\mathsf{S}
$, and the q-capacity $\mathsf{C}_{q}\left( \Lambda\right) $ is equal $%
\mathsf{C}_{\mathcal{B}^{1}}=\ln\dim \mathcal{B}^{1}$.%
\endproof%
%

In the general case, d-entanglements can be more informative than
o-entanglements as can be shown by an example of a quantum noisy channel for
which 
\begin{equation*}
\mathsf{I}_{1}\left( \varsigma_{1},\Lambda\right) >\mathsf{I}_{o}\left(
\varsigma_{1},\Lambda\right) ,\quad\mathsf{C}_{1}\left( \Lambda\right) >%
\mathsf{C}_{o}\left( \Lambda\right) .
\end{equation*}
The last equalities of the above theorem are related to the work on entropy
by Voiculescu \cite{Voi}.

\end{document}